\documentclass[parskip=half]{scrartcl}

\usepackage[english]{babel}
\usepackage[T1]{fontenc}
\usepackage[utf8]{inputenc}

\usepackage{authblk} 

\usepackage{chngcntr}		
	\counterwithout{equation}{section}			

\makeatletter 
\renewcommand{\section}{\scr@startsection%
    {section}
    {1}
    {0mm}
    {1\baselineskip}
    {0.1\baselineskip}
    {\Large\sffamily\bfseries}}
    
\renewcommand{\subsection}{\scr@startsection%
    {subsection}
    {2}
    {0mm}
    {0.5\baselineskip}
    {0.1\baselineskip}
    {\large\sffamily\bfseries}}
\makeatother 

\newcommand{\boxeq}[1]{%
    \fbox{\parbox{\textwidth - 2\fboxsep - 2\fboxrule}{
    \begin{equation}
        #1
    \end{equation}
    }}{\parfillskip=0pt\par}}

\usepackage[shortlabels]{enumitem}

\usepackage[labelformat=simple]{subcaption}
\usepackage{amssymb}
\usepackage{amsmath}
\usepackage{mathtools}
    \mathtoolsset{showonlyrefs=true}

\usepackage{bm}         
    
\usepackage{array}
    \newcolumntype{L}{>{$\displaystyle}l<{$}}
    
\usepackage{diagbox}
    
\usepackage[dvipsnames, table, xcdraw]{xcolor}

\usepackage{currfile}   

\usepackage{tikz}
    \usetikzlibrary{calc}   
    \usetikzlibrary{fadings}    
        \newenvironment{externalize}{}{}
        
\usepackage{pgfplots}
    \pgfplotsset{compat=1.16}
    \usepgfplotslibrary{fillbetween}
    \usepackage{pgfmath}
    \usepackage{pgfplotstable}
 
\usepackage{xifthen}        
\usepackage{xstring}        

\usepackage{hyperref}   
    \hypersetup{colorlinks=true, linkcolor=blue, citecolor=blue, urlcolor=blue}  
    
    \newcommand{\doi}[1]{DOI \href{https://doi.org/#1}{#1}}

\usepackage{orcidlink}

\usepackage[normalem]{ulem}  

\usepackage{todonotes}
    \makeatletter   
        \renewcommand{\todo}[2][]{\tikzexternaldisable\@todo[#1]{#2}\tikzexternalenable}
    \makeatother

\newcommand{\refFig}[1]{Figure~\ref{fig:#1}}

\newcommand{\refSec}[1]{Section~\ref{sec:#1}}
\newcommand{\refTable}[1]{Table~\ref{table:#1}}

\newcommand{\N}{\mathbb{N}}
\newcommand{\abs}[1]{\left\lvert#1\right\rvert}
\newcommand{\cupdot}{\mathbin{\mathaccent\cdot\cup}}
\newcommand{\tinyfrac}[2]{\raisebox{0.7pt}{$\scriptstyle \frac{#1}{#2}$}}
\newcommand{\supertinyfrac}[2]{\raisebox{0.7pt}{\text{\tiny$\scriptscriptstyle\frac{#1}{#2}$}}}
\newcommand{\tuple}[1]{(\row{#1}, \col{#1})}
\newcommand{\edgesInter}{E_{\text{inter}}}
\newcommand{\edgesCell}{E_{\text{cell}}}
\newcommand{\mex}{\mathcal{X}}
\newcommand{\mexHori}{\mex_{\text{hori}}}
\newcommand{\mexVert}{\mex_{\text{vert}}}
\newcommand{\mexMix}{\mex_{\text{mix}}}
\newcommand{\verticesHori}{V_{\text{hori}}}
\newcommand{\verticesVert}{V_{\text{vert}}}
\newcommand{\row}[1]{r_{#1}}
\newcommand{\col}[1]{c_{#1}}
\newcommand{\brokenHori}{B_{\text{hori}}}
\newcommand{\brokenVert}{B_{\text{vert}}}
\newcommand{\sR}{s_{\!\text{\tiny$R$}}}
\newcommand{\sC}{s_{\!\text{\tiny$C$}}}
\newcommand{\closure}[2][3]{{}\mkern#1mu\overline{\mkern-#1mu#2}}

\definecolor{lightgreen}{RGB}{146,208,80}
\definecolor{lightblue}{RGB}{0,176,240}
\definecolor{darkgreen}{rgb}{0,0.39,0}
\definecolor{darkblue}{rgb}{0,0,0.5}
\definecolor{lightgreenblue}{RGB}{73,192,160}
\definecolor{darkgreenblue}{rgb}{0,0.195,0.25}

\def\scale{0.35}
\def\LineWidth{9pt}

\def\rows{3}
\def\cols{3}
\def\shift{6}

\tikzstyle{node} = [circle, draw, fill=white, inner sep=0pt, text centered, minimum height=3ex*\scale]
\tikzstyle{nodec} = [circle, draw, fill=lightgreen, inner sep=0pt, text centered, minimum height=3.5ex*\scale]
\tikzstyle{broken} = [circle, fill=white, densely dashed, inner sep=0pt, text centered, minimum height=3ex*\scale]
\tikzstyle{every text node part}=[font=\scriptsize]

\newcommand{\setChimeraCoordinates}{
	\foreach \r in {1,...,\rows}{
		\foreach \c in {1,...,\cols}{
			\coordinate (\r-\c-h1) 	at ($(0 + \shift*\c,-2 - \shift*\r) + (-\shift,\shift+2)$);
			\coordinate (\r-\c-h2) 	at ($(1.333 + \shift*\c,-2 - \shift*\r) + (-\shift,\shift+2)$);
			\coordinate (\r-\c-h3) 	at ($(2.666 + \shift*\c,-2 - \shift*\r) + (-\shift,\shift+2)$);
			\coordinate (\r-\c-h4) 	at ($(4 + \shift*\c,-2 - \shift*\r) + (-\shift,\shift+2)$);
			
			\coordinate (\r-\c-v1) 	at ($(2 + \shift*\c, 0 - \shift*\r) + (-\shift,\shift+2)$);
			\coordinate (\r-\c-v2) 	at ($(2 + \shift*\c,-1.333 - \shift*\r) + (-\shift,\shift+2)$);
			\coordinate (\r-\c-v3) 	at ($(2 + \shift*\c,-2.666 - \shift*\r) + (-\shift,\shift+2)$);
			\coordinate (\r-\c-v4) 	at ($(2 + \shift*\c,-4 - \shift*\r) + (-\shift,\shift+2)$);
	}}
}

\newcounter{drawChimeraCounter}
\newcommand{\drawChimera}{

	\setChimeraCoordinates

	\foreach \r in {1,...,\rows}{
		\foreach \c in {1,...,\cols}{
			
			\foreach \p in {1,...,4}{
				\node[node] (\r-\c-h\p) at (\r-\c-h\p) {};
				\node[node] (\r-\c-v\p) at (\r-\c-v\p) {};
			}
			
			\foreach \x in {1,2,3,4}{
				\foreach \y in {1,2,3,4}{
					\draw (\r-\c-h\x) -- (\r-\c-v\y);
			}}
	}}
	
	\foreach \r in {1,...,\rows} {
		\foreach \c in {2,...,\cols} {
			\setcounter{drawChimeraCounter}{\c - 1}
			\foreach \x in {1,2,3,4}{
				\draw (\r-\thedrawChimeraCounter-v\x) -- (\r-\c-v\x);
		}
	}}

	\foreach \c in {1,...,\cols} {
		\foreach \r in {2,...,\rows} {
			\setcounter{drawChimeraCounter}{\r - 1}
			\foreach \x in {1,2,3,4}{
				\draw (\thedrawChimeraCounter-\c-h\x) -- (\r-\c-h\x);
		}
	}}

}

\newcommand{\clipChimera}[4]{
	\coordinate (m1) at ($0.5*(#1-#2-v1)+0.5*(#1-#2-h1)$);
	\coordinate (m2) at ($0.5*(#3-#4-v4)+0.5*(#3-#4-h4)$);
	\clip ($(m1)+(-2,+2)$) rectangle ($(m2)+(2,-2)$);
}

\newtest{\isEdgeBroken}[2]{
	\isQubitBroken{#1} \or \isQubitBroken{#2}
}

\def\brokenQubitsColor{Gray}
\newboolean{drawBrokenQubits}
\setboolean{drawBrokenQubits}{true}
\newcounter{drawBrokenChimeraCounter}
\newcommand{\drawBrokenChimera}{

	\setChimeraCoordinates
	
	\foreach \r in {1,...,\rows}{
	\foreach \c in {1,...,\cols}{
		
		\foreach \p in {1,...,4}{
			\drawQubit{\r}{\c}{h\p}
			\drawQubit{\r}{\c}{v\p}
		}
		
		\foreach \x in {1,2,3,4}{
			\foreach \y in {1,2,3,4}{
				\drawEdgeIntra{\r}{\c}{h\x}{v\y}
		}}
	}}
	
	\foreach \r in {1,...,\rows} {
	\foreach \c in {2,...,\cols} {
			\setcounter{drawBrokenChimeraCounter}{\c - 1}
			\foreach \x in {1,2,3,4}{
				\drawEdge{\r}{\thedrawBrokenChimeraCounter}{v\x}{\r}{\c}{v\x}
		}
	}}

	\foreach \c in {1,...,\cols} {
		\foreach \r in {2,...,\rows} {
			\setcounter{drawBrokenChimeraCounter}{\r - 1}
			\foreach \x in {1,2,3,4}{
				\drawEdge{\thedrawBrokenChimeraCounter}{\c}{h\x}{\r}{\c}{h\x}
		}
	}}
}

\newcommand{\drawQubit}[3]{
	\ifthenelse{\isQubitBroken{(#1-#2-#3)}}{
		\ifthenelse{\boolean{drawBrokenQubits}}{
			\node[broken, draw=\brokenQubitsColor] (#1-#2-#3) at (#1-#2-#3) {};}{}
	}{
		\node[node] (#1-#2-#3) at (#1-#2-#3) {};
	}
}

\newcommand{\drawEdgeIntra}[4]{
	\drawEdge{#1}{#2}{#3}{#1}{#2}{#4}
}

\newcommand{\drawEdge}[6]{
	\ifthenelse{\isEdgeBroken{(#1-#2-#3)}{(#4-#5-#6)}}{
		\ifthenelse{\boolean{drawBrokenQubits}}{		
			\draw[\brokenQubitsColor, dashed] (#1-#2-#3) -- (#4-#5-#6);}{}
	}{
		\draw (#1-#2-#3) -- (#4-#5-#6);
	}
}

\newcommand{\setcolor}[1]{
	\IfEqCase{#1}{%
		{1}{\colorlet{currentcolor}{lightblue}}%
		{2}{\colorlet{currentcolor}{NavyBlue}}%
		{3}{\colorlet{currentcolor}{Blue}}%
		{4}{\colorlet{currentcolor}{darkblue}}%
		{5}{\colorlet{currentcolor}{darkgreen}}%
		{6}{\colorlet{currentcolor}{ForestGreen}}%
		{7}{\colorlet{currentcolor}{lightgreen}}%
		{8}{\colorlet{currentcolor}{GreenYellow}}%
		{9}{\colorlet{currentcolor}{orange}}%
		{10}{\colorlet{currentcolor}{red}}%
		{11}{\colorlet{currentcolor}{purple}}%
		{12}{\colorlet{currentcolor}{violet}}%
	}[\colorlet{currentcolor}{#1}]%
}

\newcounter{colorCounter}
\newcommand{\setcolorCounter}{
	\setcolor{\thecolorCounter}
	\stepcounter{colorCounter}
	\ifthenelse{\thecolorCounter > 12}{\setcounter{colorCounter}{1}}{}
}

\newcounter{drawCrossFromToCounter}
\newcommand{\drawCrossFromTo}[9]{

	\draw[line width=\scale*\LineWidth, #9] (#1-#2-#3) -- (#1-#2-#4);
	
 	\foreach \x in {#5,...,#6} {
 		\ifthenelse{\x < #6}{
 			\setcounter{drawCrossFromToCounter}{\x +1}
 			\drawColoredEdge{#1}{\thedrawCrossFromToCounter}{#3}{#1}{\x}{#3}{#9}
 		}{}
 		\node[nodec, fill = #9] at (#1-\x-#3) {};
 	}
 	\foreach \x in {#7,...,#8} {
 		\ifthenelse{\x < #8}{
 			\setcounter{drawCrossFromToCounter}{\x +1}
 			\drawColoredEdge{\thedrawCrossFromToCounter}{#2}{#4}{\x}{#2}{#4}{#9}
 		}{}
 		\node[nodec, fill = #9] at (\x-#2-#4) {};
 	}
}

\newcommand{\drawCrossFromToCC}[8]{
	\setcolorCounter
	\drawCrossFromTo{#1}{#2}{#3}{#4}{#5}{#6}{#7}{#8}{currentcolor}
}

\newcommand{\drawColoredEdge}[7]{
	\draw[line width=\scale*\LineWidth, #7] (#1-#2-#3) -- (#4-#5-#6);
}

\newcommand{\drawCrossLow}[5]{
	\drawCrossFromTo{#1}{#2}{#3}{#4}{1}{#1}{#2}{\rows}{#5}
}

\newcommand{\drawCrossLowCC}[4]{
	\setcolorCounter
	\drawCrossLow{#1}{#2}{#3}{#4}{currentcolor}
}

\newcommand{\drawCross}[5]{
	\drawCrossFromTo{#1}{#2}{#3}{#4}{1}{\cols}{1}{\rows}{#5}
}

\newcommand{\drawCrossCC}[4]{
	\setcolorCounter
	\drawCross{#1}{#2}{#3}{#4}{currentcolor}
}

\newcommand{\drawCrossroad}[5]{
	\drawCrossFromTo{#1}{#2}{#3}{#4}{#2}{#2}{#1}{#1}{#5}
}

\newcommand{\drawCrossroadCC}[4]{
	\setcolorCounter
	\drawCrossroad{#1}{#2}{#3}{#4}{currentcolor}
}

\newtest{\isVeryBroken}[1]{
	\equal{#1}{(1-1-h2)}
		\OR \equal{#1}{(1-1-h3)}
		\OR \equal{#1}{(1-1-h4)}
		\OR \equal{#1}{(1-1-v3)}
		\OR \equal{#1}{(1-1-v4)}
		\OR \equal{#1}{(1-2-h1)}
		\OR \equal{#1}{(1-2-h3)}
		\OR \equal{#1}{(1-2-h4)}
		\OR \equal{#1}{(1-2-v3)}
		\OR \equal{#1}{(1-2-v4)}
		\OR \equal{#1}{(1-3-h3)}
		\OR \equal{#1}{(1-3-h4)}
		\OR \equal{#1}{(1-3-v3)}
		\OR \equal{#1}{(1-3-v4)}
		\OR \equal{#1}{(2-1-h2)}
		\OR \equal{#1}{(2-1-h3)}
		\OR \equal{#1}{(2-1-h4)}
		\OR \equal{#1}{(2-1-v1)}
		\OR \equal{#1}{(2-1-v2)}
		\OR \equal{#1}{(2-1-v3)}
		\OR \equal{#1}{(2-1-v4)}
		\OR \equal{#1}{(2-2-h3)}
		\OR \equal{#1}{(2-2-h4)}
		\OR \equal{#1}{(2-2-v2)}
		\OR \equal{#1}{(2-2-v3)}
		\OR \equal{#1}{(2-2-v4)}
		\OR \equal{#1}{(2-3-h3)}
		\OR \equal{#1}{(2-3-h4)}
		\OR \equal{#1}{(2-3-v1)}
		\OR \equal{#1}{(2-3-v2)}
		\OR \equal{#1}{(2-3-v3)}
		\OR \equal{#1}{(2-3-v4)}
		\OR \equal{#1}{(3-1-h3)}
		\OR \equal{#1}{(3-1-h4)}
		\OR \equal{#1}{(3-1-v3)}
		\OR \equal{#1}{(3-1-v4)}
		\OR \equal{#1}{(3-2-h2)}
		\OR \equal{#1}{(3-2-h3)}
		\OR \equal{#1}{(3-2-h4)}
		\OR \equal{#1}{(3-2-v3)}
		\OR \equal{#1}{(3-2-v4)}
		\OR \equal{#1}{(3-3-h3)}
		\OR \equal{#1}{(3-3-h4)}
		\OR \equal{#1}{(3-3-v1)}
		\OR \equal{#1}{(3-3-v3)}
		\OR \equal{#1}{(3-3-v4)}		
}

\begin{document}

\newbox{\orcidboxelli}
\sbox{\orcidboxelli}{\orcidlink{0000-0002-3473-8906}}
\newcommand{\orcidelli}{\usebox{\orcidboxelli}}

\newbox{\orcidboxlukas}
\sbox{\orcidboxlukas}{\orcidlink{0000-0003-4180-0744}}
\newcommand{\orcidlukas}{\usebox{\orcidboxlukas}}

\newbox{\orcidboxtobias}
\sbox{\orcidboxtobias}{\orcidlink{0000-0001-5445-8082}}
\newcommand{\orcidtobias}{\usebox{\orcidboxtobias}}

\title{Embedding of Complete Graphs in Broken Chimera Graphs}
\author[1]{Elisabeth Lobe \orcidelli}
\author[2]{Lukas Schürmann \orcidlukas}
\author[1]{Tobias Stollenwerk \orcidtobias} 
\affil[1]{German Aerospace Center (DLR), Institute for Software Technology, Germany}
\affil[2]{University Bonn, Institute of Computer Science, Germany}

\date{\normalsize\today}

\maketitle

\paragraph{Abstract} 

In order to solve real world combinatorial optimization problems with a \mbox{D-Wave} quantum annealer 
it is necessary to embed the problem at hand into the \mbox{D-Wave} hardware graph, namely Chimera or Pegasus.
Most hard real world problems exhibit a strong connectivity. 
For the worst case scenario of a complete graph, there exists an efficient solution for the embedding into the ideal Chimera graph.
However, since real machines almost always have broken qubits it is necessary to find an embedding into the broken hardware graph.

We present a new approach to the problem of embedding complete graphs into broken Chimera graphs.
This problem can be formulated as an optimization problem, more precisely as a matching problem with additional linear constraints. 
Although being NP-hard in general it is fixed parameter tractable in the number of inaccessible vertices in the Chimera graph.
We tested our exact approach on various instances of broken hardware graphs, both related to real hardware as well as randomly generated.
For fixed runtime, we were able to embed larger complete graphs compared to previous, heuristic approaches.
As an extension, we developed a fast heuristic algorithm which enables us to solve even larger instances.
We compared the performance of our heuristic and exact approaches.

\def\and{, }
\paragraph{Keywords}  
Graph minor embedding\and 
Chimera\and 
integer linear programming\and
bipartite matching\and
quantum annealing



\section{Introduction}\label{sec:intro}

\subsection{Background}\label{sec:background}

Quantum Annealing is a promising new technology which gained attention in recent years 
due to the development of commercial quantum annealing devices by the company D-Wave Systems.
These machines sample from the low energy distribution of a tunable system of interacting quantum bits (\emph{qubits})~\cite{dwave2020technical}.
The energy of the system can be described by an \emph{Ising model} including local energy fields for single qubits and certain pairwise interactions between the qubits.
However, not every qubit interacts with all other qubits.
The available couplings can be represented as edges in a graph where every qubit corresponds to a vertex.
For currently operating D-Wave hardware these graphs are the so-called \emph{Chimera} and \emph{Pegasus} graph~\cite{boothby2020next}.

In practice, no ideal Chimera or Pegasus graphs can be realized. 
Usually there are some qubits or rarely couplings which are taken offline because they do not behave as expected after calibration.
In the following we refer to the corresponding vertices as \emph{broken vertices} and to graphs containing them as \emph{broken graphs}.
Since calibrations are repeated on the order of months or years, a broken hardware graph is of practical relevance for the same amount of time.

Typical applications, however, need much more couplings than typical hardware graphs provide~\cite{venturelli2015quantum,rieffel2015case,stollenwerk2019quantum,stollenwerk2019flight}.
This problem can be mitigated by a so-called \emph{embedding}: 
One vertex of the original graph, also referred to as \emph{logical vertex}, is mapped to several qubits, also called \emph{physical} vertices, 
of the hardware graph such that the induced subgraph is connected. 
For each edge in the original graph there needs to exist at least one edge connecting the corresponding subgraphs of the two concerned logical vertices.  
See e.g.~\cite{choi2011minor} for more details. 
Each set of physical qubits representing a single logical vertex is grouped together by coupling them strongly. 

In general, given two arbitrary graphs $G$ and $H$, to decide whether $G$ can be embedded into $H$ is NP-hard. 
It is unclear but assumed that this still holds if we fix $H$ to the broken Chimera graph.
For a few well structured graphs, like the complete or the complete bipartite graph, 
the problem is trivial if $H$ is an ideal Chimera graph, due to its regular lattice structure.
Although it is the worst case scenario having a complete graph to be embedded, 
it allows the efficient embedding of all subgraphs of the complete graph.
However, if there are just a few inconveniently placed broken vertices the scheme for the ideal Chimera cannot be applied.
An important question of practical relevance is therefore:
Given a broken Chimera graph, what is the largest complete graph that can be embedded? 
This is an optimization problem we refer to in the following as \emph{largest complete graph embedding} (LCGE) problem. 

If a graph is embeddable into another it is a so-called \emph{minor} of the second graph. 
Therefore the LCGE problem is equivalent to the search for the \emph{largest clique minor}~\cite{boothby2016fast}, 
as the complete graph can be supplemented by the vertices which are not used for the embedding of the complete graph, forming a larger graph.  
The largest clique of this minor corresponds to the maximal embeddable complete graph.  

In this work we focus on the Chimera graph and leave the extension to the Pegasus graph, which has a larger connectivity for the same number of vertices~\cite{boothby2020next}, to future research.
As Pegasus is derived from the Chimera we are confident that our results are transferable. 

\subsection{Related Work}\label{sec:related}

Graph minors have been a research topic of high interest even before the D-Wave machine was released. 
Especially the work of Robertson and Seymour has mainly influenced the developments in this area. 
For instance in~\cite{robertson1995graph} they show among others that for every fixed graph $G$, 
there is a polynomial algorithm to decide whether graph $G$ is a minor of $H$ for some input graph $H$. 
For the reverse case, as we deal with here, there are no comparable results known, even for such a well structured graph as the Chimera.
Nevertheless, as the embedding is the first step to be able to run experiments on the D-Wave quantum annealing machine it is studied broadly in this context. 

Apart from problem specific approaches, as e.g.\ in~\cite{rieffel2015case}, 
current research mainly splits up into two directions: 
On the one hand the goal is to develop an efficient generic heuristic that can embed as many graphs as possible. 
The first polynomial algorithm was shown by Cai et al.\ in~\cite{cai2014practical}
and is based on finding shortest paths in the hardware graph $H$. 
As it considers both, $G$ and $H$, to be arbitrary input graphs, broken vertices in a non-ideal Chimera are already taken into account. 
It is still the standard algorithm the package \texttt{minorminor} of the D-Wave API is based on~\cite{minorminor2020repo}. 
An improvement of this algorithm is suggested in~\cite{pinilla2019layout}
and just recently compared to two new algorithms of Zbinden et al.~\cite{zbinden2020embedding}, which show even better performance. 

Those heuristic approaches work well in practice especially for less broken Chimera graphs or sparse input graphs. 
However, they have a drawback: 
If the heuristic fails to embed a graph it remains unclear if an embedding is not possible at all or the heuristic just could not find it.
There is no guarantee that an embedding can be found or how often the heuristic needs to be repeated until we find one if it exists.
Thus the second strategy is to insert an intermediate step in the embedding process
by using a template with a precomputed fixed embedding acting as a 'virtual hardware' graph.      
This template has a much simpler structure than the Chimera graph. 
Thus on the one hand the computational resources needed to calculate an embedding is decreased,
and once it is found, it can be reused for the whole operational period of the machine.   
On the other hand simple certificates can be formulated whether a graph is embeddable or not. 

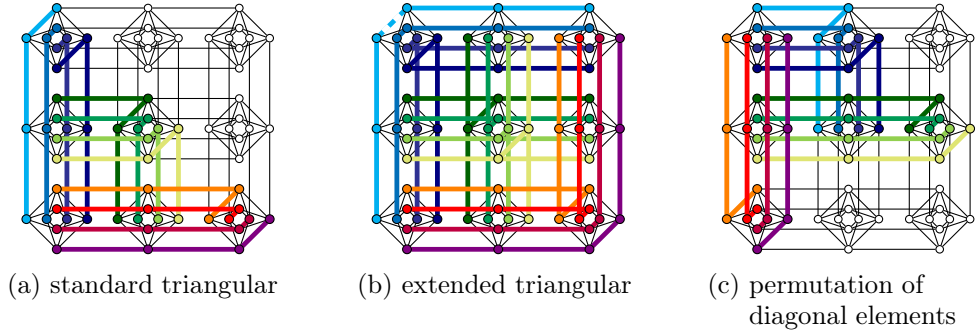
\begin{figure}[t!]
    \centering
    \def\scale{0.2}
    \captionsetup[subfloat]{singlelinecheck=off, justification=raggedright}
        \subfloat[standard triangular\\ \phantom{x}\label{fig:complete:standard}]{

\begin{externalize}
\begin{tikzpicture}[scale=\scale]

	\drawChimera
	
	\foreach \rc in {1, 2, 3}{
		\foreach \d in {1, ..., 4}{
			\drawCrossLowCC{\rc}{\rc}{v\d}{h\d}
		}
	}
	
\end{tikzpicture}
\end{externalize}} \hspace{2em}
        \subfloat[extended triangular\\ \phantom{x}\label{fig:complete:extended}]{

\begin{externalize}
\begin{tikzpicture}[scale=\scale]

	\drawChimera
	
	\foreach \rc in {1, 2, 3}{
		\foreach \d in {1, ..., 4}{
			\drawCrossCC{\rc}{\rc}{v\d}{h\d}
		}
	}
	
	\draw[line width=\scale*\LineWidth+0.2, dashed, white, shorten <=2] (1-1-v1) -- (1-1-h1);

\end{tikzpicture}
\end{externalize}}  \hspace{2em}
        \subfloat[permutation of diagonal elements\label{fig:complete:mixedupall}]{

\begin{externalize}
\begin{tikzpicture}[scale=\scale]

	\drawChimera

	\drawCrossFromToCC{1}{2}{v1}{h1}{1}{2}{1}{2}
	\drawCrossFromToCC{1}{2}{v2}{h2}{1}{2}{1}{2}
	\drawCrossFromToCC{1}{2}{v3}{h3}{1}{2}{1}{2}
	\drawCrossFromToCC{1}{2}{v4}{h4}{1}{2}{1}{2}

 	\drawCrossFromToCC{2}{3}{v1}{h1}{1}{3}{2}{2}
	\drawCrossFromToCC{2}{3}{v2}{h2}{1}{3}{2}{2}
	\drawCrossFromToCC{2}{3}{v3}{h3}{1}{3}{2}{2}
	\drawCrossFromToCC{2}{3}{v4}{h4}{1}{3}{2}{2}

	\drawCrossFromToCC{3}{1}{v1}{h1}{1}{1}{1}{3}
	\drawCrossFromToCC{3}{1}{v2}{h2}{1}{1}{1}{3}
	\drawCrossFromToCC{3}{1}{v3}{h3}{1}{1}{1}{3}
	\drawCrossFromToCC{3}{1}{v4}{h4}{1}{1}{1}{3}

\end{tikzpicture}
\end{externalize}} 
    \caption{Different versions of complete graph embeddings in ideal Chimera graph. Each color represents a single logical vertex.}
    \label{fig:complete}
\end{figure}

A universal template is the complete graph enabling to embed all graphs with the same or a smaller number of vertices or edges. 
Due to physical restrictions the Chimera graph of D-Wave was designed to be sparse but nevertheless yield an efficient embedding of the complete graph~\cite{choi2011minor}.  
The TRIAD layout, presented in~\cite{choi2011minor}, forms the basis for the triangle embedding structure of the complete graph in the (ideal) Chimera graph as shown in \refFig{complete:standard} for $K_{12}$.  
There the set of qubits representing single logical vertex forms a so-called \emph{chain}.

By extending the triangle structure each of the chains becomes cross-shaped, therefore we call them \emph{crosses} in the following.  
Additionally each pair of crosses is now connected by two edges. 
Due to this redundancy the embedding can be extended by splitting one of the crosses into its vertical and horizontal part. 
Thus we gained one additional logical vertex, as can be seen in \refFig{complete:extended}.

Unfortunately, due to broken physical vertices these schemes are unlikely to be applicable in real hardware.  
Thus in~\cite{klymko2014adiabatic} an algorithm was proposed trying to extract a subgraph of the Chimera 
where the extended triangle embedding can still be applied and has as many chains as possible. 
In~\cite{boothby2016fast} this approach is generalized by breaking up the triangle structure 
and placing L-shaped blocks such that all of them overlap pairwise.  
The principle is illustrated in \refFig{complete:mixedupall}.
As K. Boothby is one of main contributors of the D-Wave API, we assume this algorithm is implemented in the package \texttt{minorminor} to find complete graphs. 

Due to the very limited size of the maximal complete graph there are various other graphs with less connectivity but a larger number of vertices considered, too.  
Another good template candidate is the complete bipartite graph, whose embedding is closely related to the one of the complete graph. 
The idea of~\cite{goodrich2018optimizing} is to find the smallest number of vertices that have to be split up into the two partitions  
such that the resulting graph is bipartite and thus can be embedded using this template. 
In~\cite{serra2019template} this approach is elaborated and generalized to related partitioned graph structures.

Known minors can then be collected in a lookup table. 
The authors of~\cite{hamilton2017identifying} suggest to precompute all 'maximal minors' of the complete bipartite graph. 
This means an input graph is embeddable if it is a subgraph of one of the contained minors. 
Another template family are for instance the Cartesian products of complete graphs as discussed in~\cite{zaribafiyan2017systematic}.

\subsection{Our Approach}\label{sec:approach}

The approach of Boothby et.~al.~\cite{boothby2016fast} to solve the LCGE problem shows a significant advantage over~\cite{klymko2014adiabatic} regarding graph sizes. 
In this work, we further generalize both approaches to still allow for crosses of qubits representing a single logical vertex 
but also open up the triangle structure.

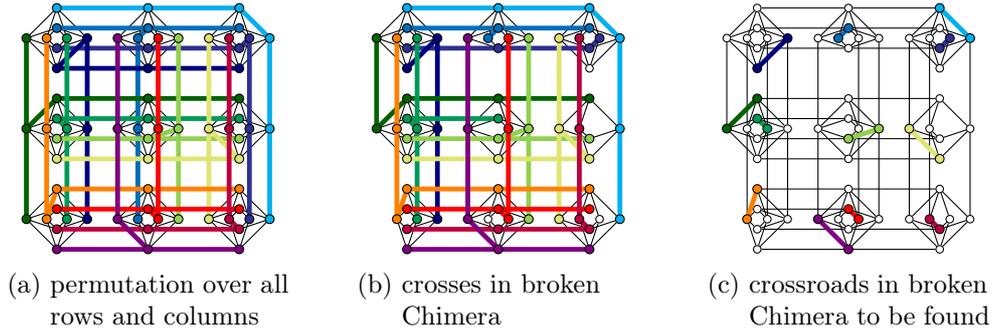
\begin{figure}[t!]
    \centering
    \def\scale{0.2}
    \captionsetup[subfloat]{singlelinecheck=off, justification=raggedright}
        \subfloat[permutation over all rows and columns\label{fig:complete:allmixed}]{

\begin{externalize}
\begin{tikzpicture}[scale=\scale]
	
	\drawChimera
	
	\drawCrossCC{1}{3}{v1}{h4}
	\drawCrossCC{1}{2}{v2}{h2}
	\drawCrossCC{1}{3}{v3}{h3}
	\drawCrossCC{1}{1}{v4}{h4}

	\drawCrossCC{2}{1}{v1}{h1}
	\drawCrossCC{2}{1}{v2}{h3}
	\drawCrossCC{2}{2}{v3}{h4}
	\drawCrossCC{2}{3}{v4}{h1}

	\drawCrossCC{3}{1}{v1}{h2}
	\drawCrossCC{3}{2}{v2}{h3}
	\drawCrossCC{3}{3}{v3}{h2}
	\drawCrossCC{3}{2}{v4}{h1}

\end{tikzpicture}
\end{externalize}} \hspace{2em}
        \subfloat[crosses in broken Chimera\label{fig:complete:crosses}]{

\begin{externalize}
\begin{tikzpicture}[scale=\scale]
	
	\setboolean{drawBrokenQubits}{false}
	
	\newtest{\isQubitBroken}[1]{
		\equal{#1}{(1-2-v4)} 
			\OR \equal{#1}{(3-1-h1)}
			\OR \equal{#1}{(2-2-h2)}
			\OR \equal{#1}{(2-3-v2)}
			\OR \equal{#1}{(2-3-h3)}
			\OR \equal{#1}{(3-3-v2)}
	}
	
	\drawBrokenChimera
	
	\drawCrossFromToCC{1}{3}{v1}{h4}{1}{3}{1}{3}
	\drawCrossFromToCC{1}{2}{v2}{h2}{1}{3}{1}{1}
	\drawCrossFromToCC{1}{3}{v3}{h3}{1}{3}{1}{1}
	\drawCrossFromToCC{1}{1}{v4}{h4}{1}{1}{1}{3}

	\drawCrossFromToCC{2}{1}{v1}{h1}{1}{3}{1}{2}
	\drawCrossFromToCC{2}{1}{v2}{h3}{1}{2}{1}{3}
	\drawCrossFromToCC{2}{2}{v3}{h4}{1}{3}{1}{3}
	\drawCrossFromToCC{2}{3}{v4}{h1}{1}{3}{1}{3}

	\drawCrossFromToCC{3}{1}{v1}{h2}{1}{3}{1}{3}
	\drawCrossFromToCC{3}{2}{v2}{h3}{1}{3}{1}{3}
	\drawCrossFromToCC{3}{3}{v3}{h2}{1}{3}{1}{3}
	\drawCrossFromToCC{3}{2}{v4}{h1}{1}{3}{1}{3}

\end{tikzpicture}
\end{externalize}} \hspace{2em} 
        \subfloat[crossroads in broken Chimera to be found\label{fig:complete:crossroads}]{

\begin{externalize}
\begin{tikzpicture}[scale=\scale]

	\setboolean{drawBrokenQubits}{false}

	\newtest{\isQubitBroken}[1]{
		\equal{#1}{(1-2-v4)} 
			\OR \equal{#1}{(3-1-h1)}
			\OR \equal{#1}{(2-2-h2)}
			\OR \equal{#1}{(2-3-v2)}
			\OR \equal{#1}{(2-3-h3)}
			\OR \equal{#1}{(3-3-v2)}
	}
	
	\drawBrokenChimera

	\drawCrossroadCC{1}{3}{v1}{h4}
	\drawCrossroadCC{1}{2}{v2}{h2}
	\drawCrossroadCC{1}{3}{v3}{h3}
	\drawCrossroadCC{1}{1}{v4}{h4}

	\drawCrossroadCC{2}{1}{v1}{h1}
	\drawCrossroadCC{2}{1}{v2}{h3}
	\drawCrossroadCC{2}{2}{v3}{h4}
	\drawCrossroadCC{2}{3}{v4}{h1}

	\drawCrossroadCC{3}{1}{v1}{h2}
	\drawCrossroadCC{3}{2}{v2}{h3}
	\drawCrossroadCC{3}{3}{v3}{h2}
	\drawCrossroadCC{3}{2}{v4}{h1}

\end{tikzpicture}
\end{externalize}}
    \caption{By permuting the crossroads we can find complete graph embeddings in a broken Chimera graph.}
    \label{fig:crossroads}
\end{figure}

\refFig{complete:allmixed} shows a variant of a complete graph embedding in the ideal Chimera graph. 
In this construction every row of qubits is connected to a column of qubits like in the extended triangular embedding in \refFig{complete:extended}.  
But in contrast to \refFig{complete:extended} the edges connecting the horizontal and vertical cross parts do not lie on the diagonal of the Chimera but are distributed over the graph. 
We call those edges \emph{crossroads} in the following.
As each of the crosses occupies the full horizontal respectively vertical part, 
every row respectively every column of qubits belongs to a specific cross.   
For each row and column combination there is a unique crossroad connecting them. 
Thus such an embedding is defined by a matching of rows to columns.
In turn, each matching of rows to columns defines a complete graph embedding in the ideal Chimera graph.  

The redundant edges connecting two crosses would again allow for one more logical vertex by spitting one of the crosses at the crossroad.  
However, we disregard this, as the redundancy offers another opportunity:
The ends of the crosses could be cut off to make room for broken qubits as shown in \refFig{complete:crosses}.
There the remaining, shorter crosses still have an edge to every other cross thus form a complete graph embedding.
But given an arbitrary broken Chimera graph, how do we place the crosses such that this is fulfilled?

By choosing a certain edge connecting a row and a column to locate a crossroad there, the corresponding cross is well defined: 
Both the horizontal and the vertical part are extended until we reach the boundary of the Chimera graph or a broken qubit.
To place two crossroads we need to ensure the resulting crosses 'meet' each other, meaning there is at least one edge connecting both.   
Thus the LCGE problem can be reformulated as: 
How do we match rows with columns to form crossroads, like in \refFig{complete:crossroads}, such that all resulting crosses meet each other? 
In particular, if the Chimera graph is very broken, which matching results in the largest possible complete graph?  
This question is an optimization problem, whose construction we show in the following sections.
 
For simplification of the notation we show the construction for the standard symmetric form of the Chimera graph, like in current hardware.
But this approach can be extended to arbitrary dimensions. 

In \refSec{chimera} we start with introducing a certain indexing of the Chimera graph, 
followed by the actual derivation of the optimization problem formulation in \refSec{ilp}.  
At the end of this section the complete ILP is summarized.
Afterwards the problem is analysed theoretically in \refSec{analysis}. 
The results of the experiments explained in \refSec{experiments} are evaluated in \refSec{results}. 
Finally, in \refSec{conclusion} we present our conclusion.

\section{Description of the Hardware Graph}\label{sec:chimera}

In this section we present the Chimera hardware graph with a specific indexing of the graph vertices, 
being suitable for the formulation of the LCGE problem, and the variable input parameters. 

\subsection{Chimera Graph Indexing}\label{sec:indexing}

First, we introduce some general notations used throughout this work.
For some $n, m \in \N$ let $[m;n] \vcentcolon = \{m, m+1, \ldots, n\}$
be the enumeration from $m$ to $n$, 
where we have $[m;n] = \emptyset$, if $n < m$. 
For shortness we use $[n] \vcentcolon = [1, n]$ for enumerating from 1, where we say $[0] = \emptyset$. 
If a set $S$ is the disjoint union of two sets $S_1$ and $S_2$, that means $S_1 \cup S_2 = S$ and $S_1 \cap S_2 = \emptyset$, we use    
$S = S_1 \cupdot S_2$. 

\begin{figure}[t]
    \def\scale{0.2}
    \centering
    \subfloat[horizontal vertices\label{fig:index:hori}]{
\begin{externalize}
\begin{tikzpicture}[scale=\scale]

	\node at (-4.2, 4) {$h = \tuple{h}$};
	\node at (2, 4) {$1$};
	\node at (8, 4) {$2$};

	\setChimeraCoordinates		
	\clip (-5.5, -9) rectangle (11, 5);
	
	\fill[lightgray, opacity=0.25] (-1,-15) rectangle (5,5);	
	\fill[lightgreen, opacity=0.25] (-6,2.5) rectangle (17,1.5);
	\fill[lightgreen, opacity=0.25] (-6,-1/6) rectangle (17,-7/6);
	\fill[lightgreen, opacity=0.25] (-6,-4.5) rectangle (17,-3.5);
	\fill[lightgreen, opacity=0.25] (-6,-1/6-6) rectangle (17,-7/6-6);
	
	\drawChimera
	
 	\newcounter{chimeraExplainedCounter1}
 	\foreach \r in {1, ..., \rows} {
 		\foreach \x in {1, ..., 4} {
 			\setcounter{chimeraExplainedCounter1}{\r*4+\x-4}
 			\node[text ragged, anchor=west] at ($(\r-1-v\x) - (7,0)$) {$\arabic{chimeraExplainedCounter1}$};
 			\foreach \c in {1, ..., \cols} {
 				\node[nodec] at (\r-\c-v\x) {};
 			}
 		}
 	}
 		
\end{tikzpicture}
\end{externalize}} \hfill 
    \subfloat[vertical vertices\label{fig:index:vert}]{
\begin{externalize}
\begin{tikzpicture}[scale=\scale]

	\node at (-4.2, 4) {$v = \tuple{v}$};
	\node[text ragged, anchor=west] at (-5,0) {$1$};
	\node[text ragged, anchor=west] at (-5,-6) {$2$};
	
	\setChimeraCoordinates
	\clip (-5.5, -9) rectangle (11, 5);

	\fill[lightgray, opacity=0.25] (-6,-3) rectangle (17,3);
	\fill[lightblue, opacity=0.15] (-0.5, 5) rectangle (0.5, -15);
	\fill[lightblue, opacity=0.15] (2+1/6, 5) rectangle (3+1/6, -15);
	\fill[lightblue, opacity=0.15] (5.5, 5) rectangle (6.5, -15);
	\fill[lightblue, opacity=0.15] (8+1/6, 5) rectangle (9+1/6, -15);
	
	\drawChimera
	
 	\newcounter{chimeraExplainedCounter2}
 	\foreach \c in {1, ..., \cols} {
 		\foreach \x in {1, ..., 4} {
 			\setcounter{chimeraExplainedCounter2}{\c*4+\x-4}
 			\node at ($(1-\c-h\x) + (0,4)$) {$\arabic{chimeraExplainedCounter2}$};
 			\foreach \r in {1, ..., \rows} {
 				\node[nodec, fill=lightblue] at (\r-\c-h\x) {};
 			}
 		}
 	}
	
\end{tikzpicture}
\end{externalize}}  ~\hfill 
    \subfloat[edge examples $(2,8)$ and $(5,1)$\label{fig:index:edge}]{
\pgfmathsetmacro\correctDist{1.2/\scale}
\begin{externalize}
\begin{tikzpicture}[scale=\scale]

	\setChimeraCoordinates
	\clip (-5, -9) rectangle (11, 5);
	
	\fill[lightblue, opacity=0.15] (-0.5, 5) rectangle (0.5, -15);
	\fill[lightblue, opacity=0.15] (9.5, 5) rectangle (10.5, -15);
	\fill[lightgreen, opacity=0.25] (-6, 1/6) rectangle (17, 7/6);
	\fill[lightgreen, opacity=0.25] (-6,-4.5) rectangle (17,-3.5);

	\drawChimera
	
	\node at (-3, 4) {$(r,c)$};
	
	\drawColoredEdge{2}{1}{v1}{2}{1}{h1}{white}
	\drawColoredEdge{2}{1}{v1}{2}{1}{h1}{path fading=east, lightblue}
	\drawColoredEdge{2}{1}{v1}{2}{1}{h1}{path fading=west, lightgreen}
	
	\node[nodec] at (2-1-v1) {};
 	\node[nodec, fill=lightblue] at (2-1-h1) {};
 	\node at ($(1-1-h1) + (0,4)$) {$1$};
 	\node at ($(1-2-h4) + (0,4)$) {$8$};
 	
	\drawColoredEdge{1}{2}{v2}{1}{2}{h4}{white}
	\drawColoredEdge{1}{2}{v2}{1}{2}{h4}{path fading=west, lightblue}
	\drawColoredEdge{1}{2}{v2}{1}{2}{h4}{path fading=east, lightgreen}
	
	\node[nodec] at (1-2-v2) {};
 	\node[nodec, fill=lightblue] at (1-2-h4) {};
 	\node[text ragged, anchor=west] at ($(1-1-v2) - (6.5,0)$) {$2$};
 	\node[text ragged, anchor=west] at ($(2-1-v1) - (6.5,0)$) {$5$};
	
\end{tikzpicture}
\end{externalize}} 
    \caption{Specific indexing in Chimera graph.}
    \label{fig:index}
\end{figure}
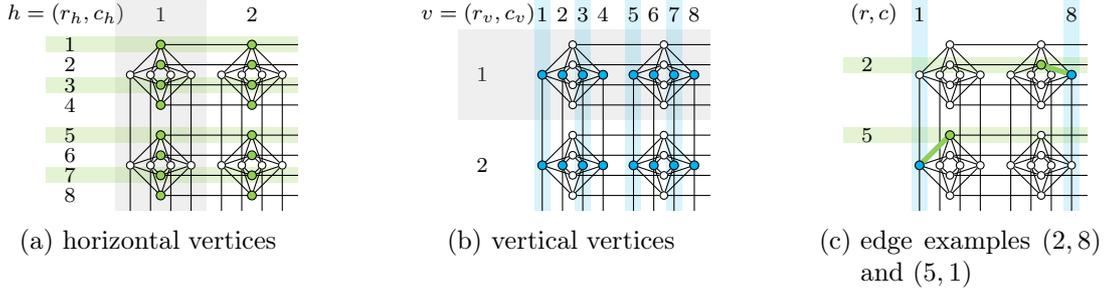

A Chimera graph is defined by a lattice structure of complete bipartite subgraphs, so called \emph{unit cells}, 
where the number of rows or columns can vary as well as the amount of vertices in the subgraph partitions. 
We refer to the latter as the \emph{depth} of the Chimera Graph.
Based on current hardware the reference is always the ideal symmetric Chimera graph
with the number of rows and columns of unit cells given by size $s \in \N$ and a depth of 4, which we denote by $C_{s,s,4}$.
Due to the lattice structure each vertex is represented as a tuple of indices referring to its row and column.
For $C_{s,s,4} = (\verticesHori \cupdot \verticesVert, \edgesCell \cupdot \edgesInter)$, 
using the index sets $S \vcentcolon= [s]$ and $N \vcentcolon= [n]$ with $n \vcentcolon = 4s$, 
we define   
\begin{itemize}
    \item the horizontally connected vertices
	    \begin{equation}
			\verticesHori \vcentcolon = N \times S
		\end{equation}
		with $n$ rows and $s$ columns, 
		which are in the vertically arranged partitions of the unit cells as illustrated in green in \refFig{index:hori}, 
		\item the vertically connected vertices
		\begin{equation}
			\verticesVert \vcentcolon = S \times N
		\end{equation}
		with $s$ rows and $n$ columns, the horizontal partition illustrated in blue in \refFig{index:vert}, 
	\item the inter unit cell edges $\edgesInter \subset \verticesHori^2 \cup \verticesVert^2$ connecting vertices of different unit cells, 
		which are not needed explicitly in the following and therefore are not precised here, and 
	\item the edges inside of the single unit cells 
		\begin{equation}
			\edgesCell \vcentcolon = \Big\{(h, v) = \big(\tuple{h}, \tuple{v}\big) \in \verticesHori \times \verticesVert \colon 
										\row{v} = u(\row{h}), \col{h} = u(\col{v})\Big\}. 
		\end{equation}
\end{itemize}

In the latter we use the function $u \colon N \to S$ with $u(x) = \left\lceil\tfrac{x}{4}\right\rceil$,
being the mapping from the \emph{inner row/column} to the \emph{unit cell row/column} index,
in the equality constraints to ensure that the paired vertices lie in the same unit cell.
Since by this the unit cell rows and columns are given implicitly we can use the congruent representation 
\begin{align}
	\edgesCell &\cong \big\{ (\row{h},  \col{v}) \colon \row{h},  \col{v} \in N \big\} \\
				&= \big\{ rc \colon r,c \in N \big\} \\
				&= N^2
\end{align}
in the following.
In general we use 
\begin{equation}
\begin{aligned}
	r_{(.)} &\colon (x_1, x_2) \mapsto x_1, \\
	c_{(.)} &\colon (x_1, x_2) \mapsto x_2
\end{aligned}
\end{equation} 
for providing the row respectively column for a given vertex, 
whereas $r$ and $c$ (without further index) always refer to some inner row respectively column indices without specifying a certain corresponding vertex.
Further we identify the tuple $(r, c)$ with the non-commutative product $rc$ for shortness to describe an inner unit cell edge.
An example for the indexing of edges can be seen in \refFig{index:edge}.

\subsection{Broken Vertices}\label{sec:broken}

With regard to real hardware we consider some vertices to be unavailable. 
For the symmetric Chimera graph $C_{s,s,4}$ with $s \in \N$ as described in the previous section 
let $\brokenHori \subset \verticesHori$ and $\brokenVert \subset \verticesVert$ be the sets of different broken vertices and $B \vcentcolon = {\brokenHori \cupdot \brokenVert}$.
In our experiments we vary the ratio of broken vertices to the total number of vertices in an ideal Chimera graph,
that is
\begin{equation}
	b \vcentcolon = \frac{\abs{B}}{\abs{\verticesHori} + \abs{\verticesVert}} = \frac{\abs{B}}{8s^2}.
\end{equation}

While for an ideal Chimera graph the set of possible crossroads defining the crosses in the embedding is just $\edgesCell$, 
the available combinations in a broken Chimera graph are restricted to those inner unit cell edges which do not contain a broken vertex:
\begin{equation}
\begin{aligned}
	A \vcentcolon ={}& \big\{ (h, v) \in \edgesCell \colon h \in \verticesHori \setminus \brokenHori, v \in \verticesVert \setminus \brokenVert\big\} \\
	 \cong{}& \big\{ rc \in N^2 \colon \big((r, u(c)), (u(r), c)\big) \in \edgesCell \cap \left(\left(\verticesHori \setminus \brokenHori\right) \times \left(\verticesVert \setminus \brokenVert\right)\right) \big\}.
\end{aligned}
\end{equation}

\section{ILP Formulation}\label{sec:ilp}

In this section we construct an integer linear optimization program (ILP) for the LCGE problem 
over arbitrary input parameters $s$, $\brokenHori$ and $\brokenVert$ as described in the previous section.

\subsection{Bipartite Matching Problem}\label{sec:matching}

In general the LGCE as we consider it here is a matching problem: 
Which row can be matched with which column to form a crossroad in an optimal embedding following our construction rules?


The decision which of the available combinations is taken can be encoded in binary problem variables $x \in \{0,1\}^A$ with
\begin{equation}
	x_{rc} = \begin{cases}
		1, &\text{if row $r$ is matched to column $c$,}\\
		0, &\text{otherwise.}
	\end{cases}
\end{equation}
We say a crossroad $rc$ is \emph{activated} if its corresponding binary variable $x_{rc}$ is activated in an optimal solution meaning it is set to 1.  
For simplification we use $x \in \{0,1\}^{N \times N}$ in the following with disabling all unavailable row column pairs by presetting $x_{rc} = 0$ for all $rc \in N^2 \setminus A$, 
although this extends the model with additional variables.

As the goal is to match as much as possible we want to maximize the number of activated binary variables, 
hence the objective is 
\begin{equation}
	\sum_{rc \in A} x_{rc} = \sum_{rc \in N^2} x_{rc}.
\end{equation}

Our construction is based on crossroads joining full rows and columns. 
Therefore only one crossroad per row and column is allowed. 
This can be enforced by the matching constraints
\begin{equation} \label{eq:matching}
\begin{aligned}
	\sum_{r\tilde{c} \in A} x_{r\tilde{c}} = \sum_{r \in N} x_{r\tilde{c}} &\leq 1 \quad\forall \tilde{c} \in N,\\
	\sum_{\tilde{r}c \in A} x_{\tilde{r}c} = \sum_{c \in N} x_{\tilde{r}c} &\leq 1 \quad\forall \tilde{r} \in N.
\end{aligned}
\end{equation}
Those types of constraints are also called \emph{cardinality constraints} 
as they enforce choosing a certain number of members, here just one, out of a given set.  
By these restrictions the optimal value of the objective function corresponds to the size, meaning the number of vertices, of the largest embeddable complete graph.
Additionally they confirm the upper bound on the objective function of $n = 4s$, 
which is the maximal size of a complete graph in $C_{s, s, 4}$ using our construction as explained in \refSec{approach}. 
	
Until now the constructed problem is just a simple \emph{maximum bipartite matching problem}. 
In the following we show further constraints that need to be added.  


\subsection{Mutually Exclusive Sets Constraints}\label{sec:constraints}

\begin{figure}
    \def\scale{0.2}
    \centering
    \subfloat[different unit cell rows \label{fig:crosses:diff}]{
\begin{externalize}
\begin{tikzpicture}[scale=\scale] 

	\newtest{\isQubitBroken}[1]{
		\equal{#1}{(2-3-v1)} 
			\OR \equal{#1}{(3-5-v3)}
	}
	
	\def\rows{4}
	\def\cols{7}
	
	\setChimeraCoordinates	
	\clipChimera{2}{2}{3}{6}
	\drawBrokenChimera
	
	\drawCrossFromTo{2}{5}{v1}{h4}{4}{7}{1}{4}{darkblue}
	\drawCrossFromTo{3}{2}{v3}{h1}{1}{4}{1}{4}{darkgreen}

\end{tikzpicture}
\end{externalize}} \\[2ex]
    \subfloat[same unit cell row \label{fig:crosses:same}]{

\begin{externalize}
\begin{tikzpicture}[scale=\scale]
	
	\newtest{\isQubitBroken}[1]{
		\equal{#1}{(2-3-v1)} 
			\OR \equal{#1}{(2-5-v3)}
	}
	
	\def\rows{3}
	\def\cols{7}
	
	\setChimeraCoordinates
	\clipChimera{2}{2}{2}{6}
	\drawBrokenChimera
	
	\drawCrossFromTo{2}{3}{v3}{h1}{1}{4}{1}{3}{darkgreen}
	\drawCrossFromTo{2}{5}{v1}{h4}{4}{7}{1}{3}{darkblue}

\end{tikzpicture}
\end{externalize}}
    \caption{Examples of crosses that do not meet due to broken vertices (gray, dashed).}
    \label{fig:crosses}
\end{figure}
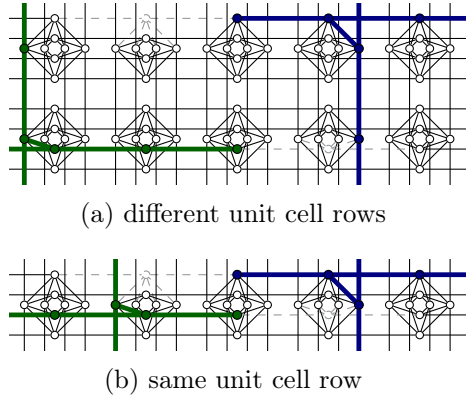  

If a horizontal vertex is broken, it interrupts the horizontal path from the left to the right.
This prevents a cross occupying this row to be extended to the boundaries of the Chimera graph. 
It is analogous for a broken vertical vertex and a cross using the corresponding column.
This needs to be taken into account when considering possible crossroad candidates for the embedding. 
\refFig{crosses} depicts an example of such a situation. 
Due to the broken vertices the corresponding crosses for certain pairs of crossroads might not meet. 
This means there do not exist any edges between the different vertices of the crosses, 
even if they reach the same unit cell like in \refFig{crosses:same}.
But at least one edge is needed to provide a valid embedding of two vertices of the complete graph. 
Therefore those crossroads cannot be activated together
and we need to introduce further constraints enforcing the activation of only one of them.

We will see that there are not only isolated pairs but clusters of crossroads all being pairwise forbidden, 
which means only one of all of them can be activated. 
We call those clusters mutually exclusive sets (MES). 
The construction of those sets differs for certain pairs of broken vertices. 
We have the following cases, which are handled separately in the next paragraphs: 
\begin{enumerate}[parsep=0pt]
    \item two broken horizontal vertices, 
    \item two broken vertical vertices,
    \item two different broken vertices, one horizontal and one vertical. 
\end{enumerate}
	
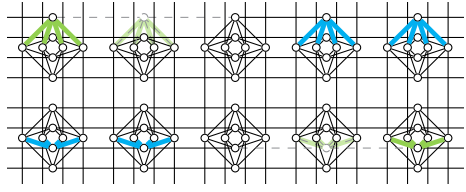
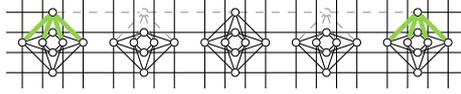
\begin{figure}
    \def\scale{0.2}
    \centering
    \subfloat[two different MES due to different rows\label{fig:mes:two}]{

\begin{externalize}
\begin{tikzpicture}[scale=\scale] 
	    
	\newtest{\isQubitBroken}[1]{
		\equal{#1}{(2-3-v1)} 
			\OR \equal{#1}{(3-5-v3)}
	}
	\def\rows{4}
	\def\cols{7}
	
	\setChimeraCoordinates	
	\clipChimera{2}{2}{3}{6}	
	
	\foreach \x in {1, ..., 4} {
		\drawColoredEdge{2}{3}{v1}{2}{3}{h\x}{lightgreen!50}
		\drawColoredEdge{3}{5}{v3}{3}{5}{h\x}{lightgreen!50}
	}
		
	\drawBrokenChimera

	\foreach \x in {1, ..., 4} {
		\drawColoredEdge{2}{2}{v1}{2}{2}{h\x}{lightgreen}
		\drawColoredEdge{3}{6}{v3}{3}{6}{h\x}{lightgreen}
		\drawColoredEdge{3}{2}{v3}{3}{2}{h\x}{lightblue}
		\drawColoredEdge{2}{6}{v1}{2}{6}{h\x}{lightblue}
		\drawColoredEdge{3}{3}{v3}{3}{3}{h\x}{lightblue}
		\drawColoredEdge{2}{5}{v1}{2}{5}{h\x}{lightblue}
	}

\end{tikzpicture}
\end{externalize}}  \\[2ex]
    \subfloat[MES in same row already handled by matching constraints\label{fig:mes:same}]{

\begin{externalize}
\begin{tikzpicture}[scale=\scale]

	\newtest{\isQubitBroken}[1]{
		\equal{#1}{(2-3-v1)} 
			\OR \equal{#1}{(2-5-v1)}
	}
	
	\def\rows{3}
	\def\cols{7}
			
	\setChimeraCoordinates
	\clipChimera{2}{2}{2}{6}
	\drawBrokenChimera
	
	\foreach \x in {1, ..., 4} {
		\drawColoredEdge{2}{2}{v1}{2}{2}{h\x}{lightgreen}
		\drawColoredEdge{2}{6}{v1}{2}{6}{h\x}{lightgreen}
	}
	
\end{tikzpicture}
\end{externalize}}
    \caption{Sets of mutually exclusive crossroads caused by two broken horizontal vertices.}
    \label{fig:mes}
\end{figure}

\paragraph{1.}
Let $h = \tuple{h} \neq k = \tuple{k} \in \brokenHori$ be \textbf{two broken horizontal vertices}, as illustrated in Figures~\ref{fig:mes:two} and~\ref{fig:mes:same}.   
Due to the horizontal interruption the crossroads on the left and the right of the two vertices are affected. 
In~\ref{fig:mes:same} both broken vertices lie in the same inner row. 
As the matching constraints already permit only one crossroad per row, no further constraints are necessary in this case
and we can restrict to vertices with $\row{h} \neq \row{k}$,  
which is the case shown in~\ref{fig:mes:two}.
There we have two MES, which are highlighted in different colors. 
Each of the blue crossroads in the right top corner cannot be activated together with the others in this corner due to the matching constraints and 
they cannot be activated together with the blue in the left bottom corner because their corresponding crosses would not meet. 
The same holds for the green crossroads in the opposite corners. 
 
For the definition of the MES we need all crossroads from the left boundary until the leftmost broken vertex and all crossroads from the rightmost broken vertex until the right boundary in the two corresponding inner rows. 
The incident edges (light green) to the broken vertices are excluded by definition of $A$, respectively set to $0$, 
but again are included in the definition of the two sets for simplicity.   
Let for example $h$ be the top left broken vertex in \refFig{mes:two} and $I_\text{left} \subseteq N$ describe the interval of columns on the left and $I_\text{right}\subseteq N$ on the right. 
The set of blue crossroads in the left top corner is then given by combining the row $\row{h}$ with each of the columns in $I_\text{left}$. 
The remaining blue crossroads combine $\row{k}$ with $I_\text{right}$. 
This results in the MES $\left(\{\row{h}\} \times I_\text{left}\right) \cup \left(\{\row{k}\} \times I_\text{right}\right)$.
For the green crossroads we get symmetrically $\left(\{\row{k}\} \times I_\text{left}\right) \cup \left(\{\row{h}\} \times I_\text{right}\right)$.

\begin{figure}
    \centering

\begin{externalize}
\begin{tikzpicture}[scale=\scale, >=stealth]
	\tikzstyle{dot}=[circle, inner sep=1pt, fill]
	
	\fill[darkgreen!50] (-9,0) rectangle (0, -7);
	\fill[darkgreen!50!lightgreen!50] (-9,0) rectangle (0, -4);
	\fill[lightgreen!50] (-9,0) rectangle (-4, -2);
	
	\fill[darkblue!50] (9,0) rectangle (0, -7);
	\fill[darkblue!50!lightblue!50] (9,0) rectangle (0, -5);
	\fill[lightblue!50] (9,0) rectangle (4, -3);
	
	\coordinate (l) at (-9, 0); 
	\coordinate (r) at (9, 0);
	\draw[->] (l) -- (r);
	
	\foreach \x in {-8, ..., 8} {
		\node[dot] at (\x, 0) {};
	}
	
	\draw (-4, 0) -- +(0, 1) node[above] (a) {$4s_1$};
	\draw (0, 0) -- +(0, 1) node[above] (b) {$4s_2$};
	\draw (4, 0) -- +(0, 1) node[above] (ta) {$4\tilde{s}_1$};
	
	\foreach \i in {1, ..., 4} {
		\coordinate (l\i) at ($(l) - (0, 2 * \i - 1)$);
		\coordinate (r\i) at ($(r) - (0, 2 * \i)$);
	}
	
	\draw[lightgreen] (l1) -- (l1 -| a);
	\node[lightgreen, anchor=east] at (l1) {$I(s_1,s_2)$};
	\foreach \x in {-8, ..., -4} {
		\node[dot, lightgreen] at (\x, -1) {};
	}
	
	\draw[lightblue] (r1) -- (r1 -| ta);
	\node[lightblue, anchor=west] at (r1) {$I(\tilde{s}_1,s_2)$};
	\foreach \x in {4, ..., 8} {
		\node[dot, lightblue] at (\x, -2) {};
	}

	\draw[darkgreen!50!lightgreen] (l2) -- (l2 -| b);
	\node[darkgreen!50!lightgreen, anchor=east] at (l2) {$I(s_2,\tilde{s}_1)$};
	\foreach \x in {-8, ..., 0} {
		\node[dot, darkgreen!50!lightgreen] at (\x, -3) {};
	}
	
	\draw[darkblue!50!lightblue] (r2) -- (r2 -| b);
	\node[darkblue!50!lightblue, anchor=west] at (r2) {$I(s_2, s_1)$};
	\foreach \x in {0, ..., 8} {
		\node[dot, darkblue!50!lightblue] at (\x, -4) {};
	}
	
	\draw[darkgreen] (l3) -- (l3 -| b);
	\node[darkgreen, anchor=east] at (l3) {$I(s_2, s_2)$};
	\foreach \x in {-8, ..., 0} {
		\node[dot, darkgreen] at (\x, -5) {};
	}
	
	\draw[darkblue] (r3) -- (r3 -| b);
	\node[darkblue, anchor=west] at (r3) {$I(s_2,s_2 - \tinyfrac{1}{2})$};
	\foreach \x in {0, ..., 8} {
		\node[dot, darkblue] at (\x, -6) {};
	}
\end{tikzpicture}
\end{externalize}
    \caption{Representation of interval function.}
    \label{fig:interval}
\end{figure}
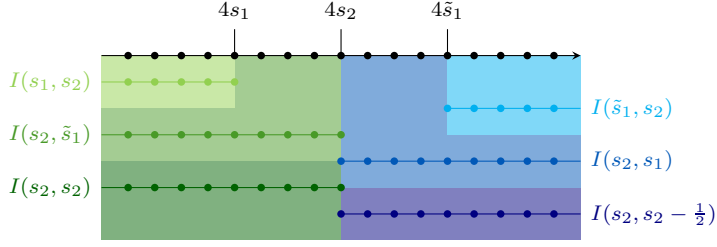

The intervals of columns, $I_\text{left}$ and $I_\text{right}$, can be derived from the broken vertices' columns depending on the relational position of the vertices.   
To describe this more formally, for the fixed size $n$ let 
\begin{equation}
	I(s_1, s_2) \vcentcolon =  \begin{cases}
		[4s_1], 	&\text{if } s_1 \leq s_2,\\
		[4(s_1 - 1) + 1; n] =  [4s_1 - 3; n], &\text{otherwise}
	\end{cases}
\end{equation}
be the interval to or from $s_1$ depending on the relation to $s_2$ for $s_1, s_2 \in S$. 
The multiplication with 4 is needed for the conversion of unit cell to inner columns. 
The behaviour of this function is illustrated in \refFig{interval} for different relations.
By the subtraction of $\tinyfrac{1}{2}$ we can circumvent the fact that $I$ only returns the left interval for identical inputs when we need the right one. 
The resulting sets of mutually exclusive crossroads can then be defined for each combination of $h$ and $k$ with    
\begin{equation}\label{eq:mes:hori}
\begin{aligned}
	X_1(h, k) &\vcentcolon = \big(\{\row{h}\} \times I(\col{h},\col{k})\big) \cup \big(\{\row{k}\} \times I(\col{k},\col{h} - \tinyfrac{1}{2})\big), \\
	X_2(h, k) &\vcentcolon = \big(\{\row{k}\} \times I(\col{h},\col{k})\big) \cup \big(\{\row{h}\} \times I(\col{k},\col{h} - \tinyfrac{1}{2})\big).
\end{aligned}
\end{equation}

Therefore we get the cardinality constraints for $i=1,2$
\begin{equation}
	\sum_{rc \in X_i(h, k)} x_{rc} \leq 1,
\end{equation}
where the sum, e.g. for $i=1$, can also be written as 
\begin{equation}
	 \sum_{c \in I(\col{h},\col{k})} x_{\row{h} c} + \sum_{c \in I(\col{k},\col{h} - \supertinyfrac{1}{2})} x_{\row{k} c}.
\end{equation}

\paragraph{2.} \textbf{Two vertical broken vertices}, with $v = \tuple{v} \neq w = \tuple{w} \in \brokenVert$, can be handled analogously to the case before by exchanging row and column: 
We can restrict on $\col{v} \neq \col{w}$. 
Taking all rows from the upper boundary to the uppermost broken vertex and all from the bottom to the lowest broken vertex results in the sets of mutually exclusive crossroads 
\begin{equation}\label{eq:mes:vert}
\begin{aligned}
	X_1(v, w) &\vcentcolon = \big(I(\row{v},\row{w}) \times \{\col{v}\}\big) \cup \big(I(\row{w},\row{v} - \tinyfrac{1}{2}) \times \{\col{w}\}\big), \\
	X_2(v, w) &\vcentcolon = \big(I(\row{v},\row{w}) \times \{\col{w}\}\big) \cup \big(I(\row{w},\row{v} - \tinyfrac{1}{2}) \times \{\col{v}\}\big).
\end{aligned}
\end{equation}
Therefore we get exemplary the constraint for $i=1$
\begin{equation}
	\sum_{rc \in X_{1}(v, w)} x_{rc} = \sum_{r \in I(\row{v},\row{w})} x_{r \col{v}} + \sum_{r \in I(\row{w},\row{v} - \supertinyfrac{1}{2})} x_{r \col{w}} \leq 1. 
\end{equation}

Let us combine the MES for all combinations in 
\begin{equation}
\begin{aligned}	
	\mexHori &\vcentcolon = \bigcup \left\{\big\{ X_1(h, k), X_2(h, k)\big\} : h, k \in \brokenHori,\, \row{h} \neq \row{k}\right\},\\
	\mexVert &\vcentcolon = \bigcup \left\{\big\{ X_1(v, w), X_2(v, w)\big\} : v, w \in \brokenVert,\, \col{v} \neq \col{w} \right\}.
\end{aligned}
\end{equation}

\begin{figure}[t!]
	\def\scale{0.2}
	\centering
	\begin{minipage}[b][9.3cm][t]{0.4\textwidth}
	   \captionsetup[sub]{margin=1em}
		\hspace{2em}\subfloat[four different relations of positions\label{fig:two:positions}]{

\begin{externalize}
\begin{tikzpicture}[scale=\scale]
	
	\newtest{\isQubitBroken}[1]{
		\equal{#1}{(3-3-v1)} 
			\OR \equal{#1}{(2-2-h4)}
			\OR \equal{#1}{(2-4-h3)}
			\OR \equal{#1}{(4-2-h1)}
			\OR \equal{#1}{(4-4-h2)}
	}
	\def\rows{5}
	\def\cols{5}
	\def\shift{8}
	
	\setChimeraCoordinates
	
	\node[Gray] at ($(3-3-v1) + (0.6, 1)$) {$(9,3)$};
	\node[Gray] at ($(2-2-h4) + (0, 4)$) {$(2,8)$};
	\node[Gray] at ($(4-2-h1) + (0, -4)$) {$(4,5)$};
	\node[Gray] at ($(4-4-h2) + (0, -4)$) {$~(4,14)$};
	\node[Gray] at ($(2-4-h3) + (0, 4)$) {$~(2,15)$};
	
	\clipChimera{2}{2}{4}{4}
	
	\drawBrokenChimera
	
	\draw[darkgreen] (3-3-v1) .. controls +(left:5) and +(south:5) .. (2-2-h4);
	\draw[darkgreen] (3-3-v1) .. controls +(right:8) and +(south:4) .. (2-4-h3);
	\draw[darkgreen] (3-3-v1) .. controls +(left:3) and +(north:2) .. (4-2-h1);
	\draw[darkgreen] (3-3-v1) .. controls +(right:4) and +(north:4) .. (4-4-h2);

	\node[darkgreen] at ($(-2, -4) + (2*\shift, -\shift)$) {Ia)};
	\node[darkgreen] at ($(-2, -4) + (2*\shift, -2*\shift)$) {IIa)};
	\node[darkgreen] at ($(-2, -4) + (3*\shift, -\shift)$) {Ib)};
	\node[darkgreen] at ($(-2, -4) + (3*\shift, -2*\shift)$) {IIb)};

\end{tikzpicture}
\end{externalize}}\\[3ex]
	\end{minipage}
	\begin{minipage}[b][9.3cm][t]{0.58\textwidth}
		\raggedleft
		\captionsetup[sub]{margin=3em}
		\subfloat[a single crossroad (green) is pairwise forbidden with all crossroads in rectangle (blue)\label{fig:two:rectangle}]{

\begin{externalize}
\begin{tikzpicture}[scale=\scale] 
	
	\tikzstyle{every text node part}=[font=\scriptsize]
	
	\newtest{\isQubitBroken}[1]{
		\equal{#1}{(4-4-v3)} 
			\OR \equal{#1}{(3-2-h1)}
	}
		      
	\def\rows{5}
	\def\cols{7}
	
	\setChimeraCoordinates	
	
	\node[Gray] at ($(4-4-v3)+(2, -3.333)$) {$\tuple{h}$};
	\node[Gray] at ($(3-2-h1)+(-3, 0)$) {$\tuple{v}$};
    \node[white] at ($(3-6-h4)+(3, 0)$) {$\tuple{v}$};
	
	\clipChimera{2}{2}{4}{6}		
	\drawBrokenChimera
	
	\drawCrossFromTo{4}{2}{v3}{h1}{2}{2}{4}{4}{lightgreen}

	\foreach \r in {2, 3} {
		\foreach \c in {4, 5, 6} {
			\foreach \x in {1, ..., 4} {
				\foreach \y in {1, ..., 4} {
					\drawColoredEdge{\r}{\c}{v\x}{\r}{\c}{h\y}{lightblue}
				}
			}
		}
	}

\end{tikzpicture}
\end{externalize}} \\[2ex]
		\subfloat[in same unit cell row it does not result in further constraints\label{fig:two:same}]{

\begin{externalize}
\begin{tikzpicture}[scale=\scale]

	\newtest{\isQubitBroken}[1]{
		\equal{#1}{(2-3-h1)} 
			\OR \equal{#1}{(2-5-v1)}
	}
	\def\rows{3}
	\def\cols{7}
	
	\setChimeraCoordinates
	
	\node[white] at ($(2-2-h4) + (0, 4)$) {$(2,8)$};		
	
    \node[white] at ($(2-2-h1)+(-3, 0)$) {$\tuple{v}$};
    \node[white] at ($(2-6-h4)+(3, 0)$) {$\tuple{v}$};
    
	\clipChimera{2}{2}{2}{6}
	
	\drawBrokenChimera

\end{tikzpicture}
\end{externalize}}
	\end{minipage}
	\captionsetup{skip=-4\baselineskip, singlelinecheck=off, justification=raggedright}
	\caption{Mutually exclusive crossroads \\ caused by two different broken \\ vertices.} 
	\label{fig:two}	
\end{figure}
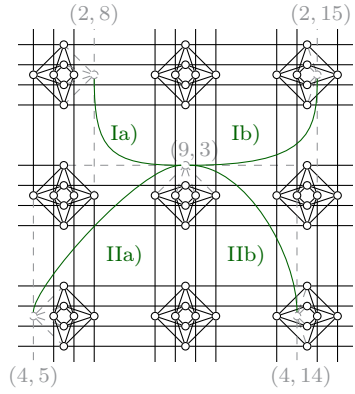
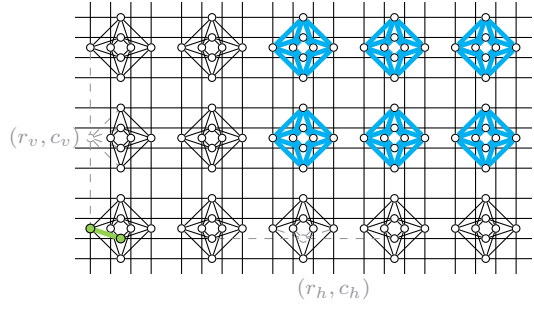
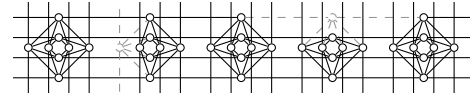

\paragraph{3.} The case with \textbf{two different broken vertices} $h = \tuple{h} \in \brokenHori$ and $v = \tuple{v} \in \brokenVert$ is different to the ones above. 
As it can be seen in \refFig{two:positions} we have four different cases depending on the relational position of the two vertices, 
whether the vertical is above or below,
\begin{enumerate}[I), parsep=0pt]
    \item $\row{v} < u(\row{h})$ or
    \item $\row{v} > u(\row{h})$,
\end{enumerate} 
and left or right, 
\begin{enumerate}[a), parsep=0pt]
    \item $\col{h} > u(\col{v})$ or
    \item $\col{h} < u(\col{v})$,
\end{enumerate}
of the horizontal broken vertex.

\refFig{two:rectangle} shows the combination \textbf{Ia)} exemplary. 
The other cases \textbf{Ib)}, \textbf{IIa)} and \textbf{IIb)} can be derived analogously but mirrored to different corners.
This is covered by the definition of $I$, which we use again in the following construction, 
therefore this holds for all cases. 
Further we can see in \refFig{two:same} that no additional constraints are provided if both vertices lie in the same unit cell row or column, respectively.
Therefore we can restrict on cases with $u(\row{h}) \neq \row{v}$ and $u(\col{v}) \neq \col{h}$.

Due to the path interruption by the broken vertices we get exactly one crossroad, $\row{h} \col{v} \cong \big(\row{h}, u(\col{v}), u(\row{h}) , \col{v}\big)$, 
in the lower left corner of \refFig{two:rectangle} illustrated in green, 
which is pairwise forbidden with all the crossroads in a rectangle, which are shown in blue.
In the following we refer to $\row{h} \col{v}$ as the \emph{common crossroad}. 
In the case shown in Ia) the rectangle includes all rows from the upper boundary until the unit cell of the broken horizontal vertex $v$ and 
all columns starting at the unit cell of the broken vertical vertex $v$ until the right boundary, 
which are the combinations in $[4\row{v}] \times [4(\col{h}-1) + 1; n]$.
More generally the rectangle is described by
\begin{equation}
	 I(\row{v}, u(\row{h})) \times I(\col{h}, u(\col{v})).
\end{equation}
The pairwise constraints
\begin{equation}
	x_{\row{h} \col{v}} + x_{rc} \leq 1 \qquad \forall rc \in I(\row{v}, u(\row{h})) \times I(\col{h}, u(\col{v}))
\end{equation}
would therefore be sufficient to describe our problem. 
But taking advantage of the matching constraints~\eqref{eq:matching} again, we can aggregate the crossroads in the rectangle: 
either all in one inner row or all in one inner column.
To keep the optimization problem description as small as possible,
we take the smallest amount of resulting MES. 
This is given by the minimum of the dimensions of the rectangle, 
hence the number of rows $\abs{I(\row{v}, u(\row{h}))}$ or the number of columns $\abs{I(\col{h}, u(\col{v}))}$. 
With
\begin{equation}
	X_{r}(h,v) \vcentcolon= \{\row{h} \col{v}\} \cup \big(\{r\} \times I(\col{h}, u(\col{v}))\big), 
\end{equation}
describing the aggregated MES for a row $r \in I(\row{v}, u(\row{h}))$, and analogously 
\begin{equation}
	X_{c}(h,v) \vcentcolon= \{\row{h} \col{v}\} \cup \big(I(\row{v}, u(\row{h})) \times \{c\}\big), 	
\end{equation}
for a column $c \in I(\col{h}, u(\col{v}))$, 
we can define
\begin{equation}
	\mexMix (h,v) \vcentcolon = \begin{cases}
		\big\{X_{r}(h,v) \colon r \in I(\row{v}, u(\row{h}))\big\}, &\text{if } \abs{I(\row{v}, u(\row{h}))} \leq \abs{I(\col{h}, u(\col{v}))}, \\
		\big\{X_{c}(h,v) \colon c \in I(\col{h}, u(\col{v}))\big\}, &\text{otherwise}, \\
	\end{cases} 
\end{equation}
choosing the set of MES with the smallest cardinality for a certain pair of broken vertices $v$ and $h$. 
With
\begin{equation}
	\mexMix \vcentcolon = \bigcup \left\{\mexMix (h,v) : (h, v) \in \brokenHori \times \brokenVert,\, \row{v} \neq u(\row{h}),\, \col{h} \neq u(\col{v})\right\}.
\end{equation} 
we can finally summarize all cardinality constraints to
\begin{equation}\label{eq:additional}
	\sum_{rc \in X} x_{rc} \leq 1 \quad \forall X \in \mexHori \cup \mexVert \cup \mexMix. 
\end{equation}


\subsection{Embedding ILP in a nutshell}\label{sec:EILP}

With the definitions of the section before we can summarize the complete embedding problem in the following ILP formulation. 
If we find an optimal solution to this ILP,
its objective value corresponds to the size of the largest embeddable complete graph 
and the activated variables define the crossroads for a corresponding embedding.     
						 		
\boxeq{\tag{E}\label{eq:EILP}
\begin{aligned}
		\max~ 		 &\sum_{rc \in N^2} x_{rc}  \\
		\text{s.t. } &\begin{aligned}[t]
							\sum_{r \in N} x_{r\tilde{c}} &\leq 1 &&\forall \tilde{c} \in N\\
							\sum_{c \in N} x_{\tilde{r}c} &\leq 1 &&\forall \tilde{r} \in N\\
							\sum_{rc \in X} x_{rc} &\leq 1 &&\forall  X \in \mexHori \cup \mexVert \cup \mexMix\\
					 		x_{rc} = 0~ &&&\forall rc \in N^2 \setminus A
						\end{aligned}\\
					 &x \in \{0,1\}^{N \times N}
\end{aligned}}

\section{Analysis}\label{sec:analysis}

In this section, we investigate the structure of the ILP Embedding problem by discussing its size, complexity and variations.
The solvability of the problem can be estimated by different parameters. 
The size of the ILP, more precisely the number of variables and constraints, is of interest
when directly passing the constructed ILP to ILP solvers and using them without any further specifications. 
However, the specific structure of the constraints allows for a deeper complexity analysis of the problem showing fixed-parameter tractability. 
At the end of this section we give a short outlook on how the ILP can be extended to more general Chimera graphs. 

\subsection{Size of the ILP}\label{sec:size}

We estimate the size of the ILP with regard to the input parameters $s$, $\brokenHori$ and $\brokenVert$.
The number of variables is $n^2 = 16s^2$ if we also take the unavailable combinations in $A$ into account. 
By removing them we get $n^2 - |A| \geq n^2 - |\brokenHori| - |\brokenVert|$, 
where the lower bound is achieved only if no two broken vertices meet in one edge.   

Apart from the $2n$ matching constraints in~\eqref{eq:matching} we show that the number of additional constraints is also polynomial in the number of broken vertices. 
We need to count the number of MES that are constructed in the former section for the different combinations of vertices. 
Taking two unequal vertices out of the broken horizontal vertices $\brokenHori$ we get
\begin{equation}
	\binom{\abs{\brokenHori}}{2} 
        = \tfrac{1}{2} \abs{\brokenHori}\left(\abs{\brokenHori}-1 \right) 
        = \tfrac{1}{2} \abs{\brokenHori}^2 - \tfrac{1}{2} \abs{\brokenHori}
\end{equation}
combinations. 
Analogously for two broken vertical vertices out of $\brokenVert$ we have 
\begin{equation}
	\binom{\abs{\brokenVert}}{2} = \tfrac{1}{2} \abs{\brokenVert}^2 - \tfrac{1}{2} \abs{\brokenVert}
\end{equation}
combinations. 
On the other hand the number of combinations for two different broken vertices is $\abs{\brokenHori}\abs{\brokenVert}$. 
Those numbers could be slightly but not significantly reduced when taking pairs into account that lie in the same rows resp. columns.

For each combination of two broken vertices of the same type we have two constraints.
Thus we have 
\begin{equation}
\begin{aligned}
    \abs{\mexHori} \leq \abs{\brokenHori}^2 - \abs{\brokenHori}\\ 
    \abs{\mexVert} \leq \abs{\brokenVert}^2 - \abs{\brokenVert}.
\end{aligned}
\end{equation}
For the different broken vertices the number of constraints depends on the size of the corresponding rectangles. 
Here we can only estimate the worst case scenario which is $n-1$ constraints, hence 
\begin{equation}
    \abs{\mexMix} \leq (n-1)\abs{\brokenHori}\abs{\brokenVert}
\end{equation} 
Therefore in total we get
\begin{equation}
	\abs{\mexHori} + \abs{\mexVert} + \abs{\mexMix} \leq \abs{\brokenHori}^2 - \abs{\brokenHori}+ \abs{\brokenVert}^2 - \abs{\brokenVert} + (n-1)\abs{\brokenHori}\abs{\brokenVert}
\end{equation}
additional cardinality constraints in~\eqref{eq:additional}.

\subsection{Problem Complexity}\label{sec:complexity}

The described problem is a matching problem on a bipartite graph. 
The simple version, without additional constraints, can be solved in polynomial time, e.g.\ with the algorithm of Hopcroft and Karp in $\mathcal{O}\left(n^{2.5}\right)$~\cite{hopcroft1973n}, 
Due to the constraints~\eqref{eq:additional}, introduced in \refSec{constraints}, our ILP corresponds to a so-called \emph{restricted maximum matching problem}.
Those problems are NP-hard in general and this even holds for cardinality constraints with a cardinality of just one like ours~\cite{tanimoto1978some}.   
But exploiting the specific structure of those constraints we can derive that the runtime is mainly dominated by the broken vertices compared to the size of the Chimera graph. 
More formally this means the problem is \emph{fixed-parameter tractable} with the number of the broken vertices $|B|$ as the parameter. 
We show the fixed-parameter tractability by enumerating the decisions that have to be made for removing constraints until the problem is a simple maximum bipartite matching problem.

Considering the constraint for two broken vertices of the same type we have two MES,~\eqref{eq:mes:hori} resp.~\eqref{eq:mes:vert}. 
As it was shown in \refSec{constraints} both MES consist of a left and a right part lying in different rows for broken horizontal vertices, 
resp.\ an upper and a lower part in different columns for broken vertical vertices. 
Since we can only take one of the crossroads in an MES into a solution, this crossroad is either in the left or in the right, resp.\ upper or lower, part. 
Imagine we decide in advance for one part of the MES.
Considering for instance some $X \in \mexHori$, with $X = \vcentcolon X_{\text{left}} \cup X_{\text{right}}$ for simplicity, 
we could choose the crossroad to be in $X_{\text{left}}$. 
Thus none of the crossroads in $X_{\text{right}}$ can be activated in the solution, 
meaning we have to set $x_{rc} = 0$ for all $rc \in X_{\text{right}}$.
The corresponding cardinality constraint reduces to  
\begin{equation}
    \sum_{rc \in X} x_{rc} = \sum_{rc \in X_{\text{left}}} x_{rc} \leq 1.
\end{equation}
As $X_{\text{left}}$ only consists of crossroads in a certain row
this constraint is weaker than the matching constraint of~\eqref{eq:matching} covering that row fully. 
Thus we can remove it and 
the resulting optimization problem, having less variables and less constraints, is easier to solve.

By considering both exclusive options, disregarding $X_{\text{right}}$ or $X_{\text{left}}$, 
and choosing the best solution we get the global optimum.
This procedure can be applied for every MES in $\mexHori$ and $\mexVert$, 
especially this can be done iteratively to already simplified versions.  
With two parts for each MES this results in total in 
\begin{equation}
	2^{\abs{\mexHori}} \cdot 2^{\abs{\mexVert}} \leq 2^{\abs{\brokenHori}^2 - \abs{\brokenHori} + \abs{\brokenVert}^2 - \abs{\brokenVert}}
\end{equation}
different simplified problems.

For different broken vertices we have much more constraints,  
but they all have one crossroad in common that cannot be matched together with the other concerned crossroads in the rectangle. 
Therefore we can proceed similarly to above. 
One option is just taking the single common crossroad into the solution and rejecting all of the rectangle.
This means the binary variable corresponding to this crossroad is set to 1 while those for the rectangle are set to 0. 
By this not only the constraints are removed but the size of the ILP is appreciably reduced, persisting for all problems resulting from subsequent decisions.
However, for an increasing size of the rectangle it gets more unlikely that the common crossroad is part of an optimal solution. 
Hence the second option is rejecting this single crossroad. 
This again results in weaker remaining constraints than the matching constraints for the whole rectangle and they can be dropped. 
These two possibilities per broken vertex pair result in total in further
\begin{equation}
	2^{\abs{\brokenHori}\abs{\brokenVert}}
\end{equation}
options.

Finally, we get at maximum
\begin{equation}
	2^{\abs{\brokenHori}\abs{\brokenVert} + \abs{\brokenHori}^2 - \abs{\brokenHori} + \abs{\brokenVert}^2 - \abs{\brokenVert}}
\end{equation}
different simplified versions of our original problem. 
As we removed all of the additional constraints along the decision tree 
they are now simple maximum bipartite matching problems and can be solved efficiently.
With 
\begin{equation}
\begin{aligned}
	&\abs{\brokenHori}^2 - \abs{\brokenHori} + \abs{\brokenVert}^2 - \abs{\brokenVert} + \abs{\brokenHori}\abs{\brokenVert} \\
	\leq{}& \abs{\brokenHori}^2 + \abs{\brokenVert}^2 + 2\abs{\brokenHori}\abs{\brokenVert}\\
	={} & \big(\abs{\brokenHori} + \abs{\brokenVert}\big)^2 = \abs{B}^2
\end{aligned}
\end{equation}
we get a worst case run time in 
\begin{equation}
	\mathcal{O}\left(2^{\abs{B}^2} \,n^{2.5}\right)
\end{equation}
and the problem is fixed parameter tractable in the choice of the broken vertices.

According to current hardware development we can reasonably assume that $|B|$ is small compared to $n$. 
Therefore, considering $|B|$ to be fixed, the problem is efficiently solvable for increasing size $n$. 
However, this means at the same time the ratio of broken vertices $R$ is decreasing as it is inversely proportional to $n^2$.  
Just keeping $R$ fixed still provides an exponential runtime. 
This aspect demands for heuristic solving approaches like presented in the following section.  
However, once the embedding is computed for a hardware graph, it can be reused during the whole operating period. 

\subsection{Generalization}\label{sec:general}

In \refSec{ilp} we have shown the construction of the ILP for the LCGE problem exemplary for the symmetric Chimera graph with a depth of 4.
But the whole setup can also be generalized for arbitrary Chimera graphs $C_{\scalebox{0.818}{\text{\normalsize$\sR$}}, \scalebox{0.818}{\text{\normalsize$\sC$}}, d}$, 
hence which are rectangular meaning $\sC \neq \sR$ or which have a different depth $d$.
In the following we briefly mention the adjustments that need to be applied. 

If $d$ is different than 4, the unit cell index function changes to $u \colon x \mapsto \left\lceil\tfrac{x}{d}\right\rceil$.
In case of $\sR \neq \sC$ we need to split $N^2$ up in $R \times C$ with the row respectively column sets $R = [d \sR]$ and $C = [d \sC]$.
Of course in this case the maximal number of vertices in an embeddable complete graph, even if it is an ideal Chimera, is just $d \min\{\sR, \sC\}$.
As the amount of rows and columns is not equal anymore we have to adjust the interval function to be able to differ between maximal row or column with
\begin{equation}
	I_{\!\text{\tiny$R/C$}}(s_1, s_2) \vcentcolon =  \begin{cases}
		[4s_1], 	&\text{if } s_1 \leq s_2,\\
		[4s_1 - 3; d s_{\!\text{\tiny$R/C$}}], &\text{otherwise}. 
	\end{cases}
\end{equation} 
With these modifications it is possible to construct the analogous matching constraints as well as the MES for the cardinality constraints. 


One might also consider to extend the model by adding broken edges, 
which could possibly be handled in an analogue case differentiation as for the broken vertices. 
But in contrast to the restriction to broken vertices the implications on the model construction and therefore the size and complexity are not trivial. 
Further a broken edge adjoining non-broken vertices is very rare and thus can be handled by marking one of the concerned vertices as broken.
Therefore we do not discuss this in more detail here.

\subsection{Delineation}\label{sec:non-optimal}

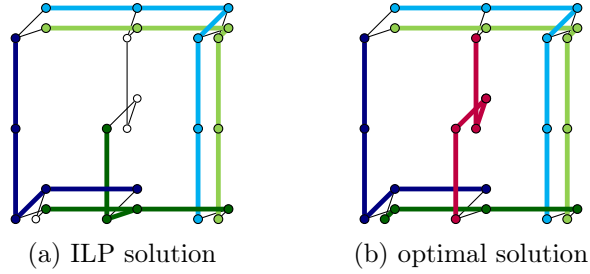
\begin{figure}
    \def\scale{0.2}
    \centering
    \subfloat[ILP solution\label{fig:noopti:ilp}]{

\begin{externalize}
\begin{tikzpicture}[scale=\scale]
	
	\setboolean{drawBrokenQubits}{false}

	\newtest{\isQubitBroken}[1]{
		\isVeryBroken{#1}
	}
	
	\drawBrokenChimera
	
 	\drawCrossFromTo{1}{3}{v2}{h2}{1}{3}{1}{3}{lightgreen}
 	\drawCrossFromTo{1}{3}{v1}{h1}{1}{3}{1}{3}{lightblue}
 	\drawCrossFromTo{3}{2}{v2}{h1}{1}{3}{2}{3}{darkgreen}
 	\drawCrossFromTo{3}{1}{v1}{h1}{1}{2}{1}{3}{darkblue}

\end{tikzpicture}
\end{externalize}} \hspace{3em}
    \subfloat[optimal solution\label{fig:noopti:opti}]{

\begin{externalize}
\begin{tikzpicture}[scale=\scale]
	
	\setboolean{drawBrokenQubits}{false}

	\newtest{\isQubitBroken}[1]{
		\isVeryBroken{#1}
	}
	
	\drawBrokenChimera
	
 	\drawCrossFromTo{1}{3}{v2}{h2}{1}{3}{1}{3}{lightgreen}
 	\drawCrossFromTo{1}{3}{v1}{h1}{1}{3}{1}{3}{lightblue}
 	\drawCrossFromTo{3}{1}{v2}{h2}{1}{3}{3}{3}{darkgreen}
 	\drawCrossFromTo{3}{1}{v1}{h1}{1}{2}{1}{3}{darkblue}
 	\drawCrossFromTo{2}{2}{v1}{h1}{2}{2}{2}{3}{purple}
 	\drawCrossFromTo{2}{2}{v1}{h2}{2}{2}{1}{2}{purple}

\end{tikzpicture}
\end{externalize}}  
    \caption{Very restricted Chimera graph containing larger complete graph than can be found with our ILP.}
    \label{fig:noopti}
\end{figure}

In our construction the embedding corresponding to a single logical vertex is formed by a cross. 
In the case of many broken vertices, this assumption might be too restrictive.
This can be seen for example in \refFig{noopti} where the solution to the ILP is not as good as the optimal solution to the LCGE problem.
We believe, however, that such corner cases are of less practical relevance since they just seem to occur for a very large ratio of broken vertices.

\section{Heuristic ILP}\label{sec:heur}

The complexity analysis in \refSec{complexity} has shown a certain structure of the additional constraints. 
Especially for the case of two broken vertices of different type there is a strong imbalance: 
Activating the single common crossroad excludes all the crossroad in the corresponding rectangle. 
Thus it is very unlikely that this crossroad is part of the optimal solution in particular for growing size of the rectangle.
In this section we show the derivation of a simpler ILP whose solution is assumed to be close to the optimal one.  

\subsection{Reducing Size}\label{sec:reduce}
 
We decided to test a heuristic approach based on excluding such unlikely common crossroads in advance. 
This reduces the number of variables and more importantly the number of constraints. 
Thus we solve only a certain part of the decision tree constructed in \refSec{complexity} 
and therefore it is not clear if the optimal value can be achieved.    

We introduce a parameter defining which common crossroads shall be removed respectively kept: 
The so called \emph{maximum rectangle ratio}, denoted here by $m$ with $0 \leq m \leq 1$,  
gives a boundary on the size of the rectangle relative to the Chimera graph size $s$ below which the common crossroad is kept.
If the number of unit cell rows times the number of columns of the rectangle exceeds $m s^2$ the crossroad is excluded.
More formally this means for two different broken vertices $h = \tuple{h} \in \brokenHori$ and $v = \tuple{v} \in \brokenVert$, that 
we do not use the crossroad $\row{h} \col{v}$, hence set $x_{\row{h} \col{v}} = 0$ in advance, if we have $M(h, v) \geq m s^2$ with 
\begin{equation}
     M(h, v) \vcentcolon =  |I(\row{v}, u(\row{h}))| \cdot |I(\col{h}, u(\col{v}))|.
\end{equation}
Thus a ratio of 1 means all common crossroads are kept while a ratio of 0 means none of them remain in the resulting optimization problem.

Given $m$, let the set of unused crossroads be
\begin{equation}
    U^m \vcentcolon = \left\{\row{h} \col{v}: (h, v) \in  \brokenHori \times \brokenVert,\,  M(h, v) \geq m s^2\right\}
\end{equation}
and further let
\begin{equation}
    \mexMix^m \vcentcolon = \bigcup \left\{\mexMix (h,v) : (h, v) \in \brokenHori \times \brokenVert,\, \row{v} \neq u(\row{h}),\, \col{h} \neq u(\col{v}),\, M(h, v) < m s^2\right\}.  
\end{equation} 
be the reduced set of MES.
With this the corresponding constraints can be simplified by replacing $\mexMix$ with $\mexMix^m$ in~\eqref{eq:EILP}.
This can be seen in the following section summarizing the heuristic ILP. 

\subsection{Heuristic Embedding ILP in a nutshell}\label{sec:HILP}

With the same definitions as before and a certain choice of $m$ we can now summarize the heuristically reduced embedding problem in the ILP formulation: 
                                
\boxeq{\tag{H}\label{eq:HILP}
\begin{aligned}
        \max~        &\sum_{rc \in N^2} x_{rc}  \\
        \text{s.t. } &\begin{aligned}[t]
                            \sum_{r \in N} x_{r\tilde{c}} &\leq 1 &&\forall \tilde{c} \in N\\
                            \sum_{c \in N} x_{\tilde{r}c} &\leq 1 &&\forall \tilde{r} \in N\\
                            \sum_{rc \in X} x_{rc} &\leq 1 &&\forall  X \in \mexHori \cup \mexVert \cup \mexMix^m\\
                            x_{rc} = 0~ &&&\forall rc \in N^2 \setminus A \cup U^m
                        \end{aligned}\\
                     &x \in \{0,1\}^{N \times N}
\end{aligned}}

\section{Experimental Setup}\label{sec:experiments}

\subsection{Random Instances}\label{sec:random}

To be able to compare our approach to current state of the art methods we consider different ratios of broken vertices for growing hardware sizes.
We have generated 10 instances for each combination of the following values:  
\begin{itemize}[parsep=0pt]
    \item sizes of Chimera graph: $s \in \{4, 6, 8, \dots 32, 34\}$,
    \item ratios of broken vertices: $b \in \{0.005, 0.01, 0.02, 0.03, 0.04, 0.05, 0.1, 0.2\}$.
\end{itemize}
The ratio of the broken vertices times the total number of vertices in the ideal Chimera graph results in the number of broken vertices for a certain size.
For each of the 10 instances we randomly chose this number out of all vertices and marked them as being broken. 
Due to rounding to whole vertices the resulting exact ratios differ slightly from the aimed ones above, especially for smaller graph sizes.

As a reference we like to remark the parameters of two real D-Wave 2000Q systems.
First, the solver \texttt{DW\_2000Q\_6}, which we accessed though the Jülich UNified Infrastructure for Quantum computing (JUNIQ),
has a size of 16 and seven broken vertices. 
This corresponds to a ratio of about 0.0034.
Second, the older USRA/NASA chip with the same size had 17 broken vertices, resulting in a ratio of about 0.0083~\cite{junger2019performance}.
Thus our experiments are much more exhaustive than current hardware demands.
 
For the heuristic approach we use $m=0$ and $m=0.25$ for our experiments as reference points  
to evaluate the impact of removing a significant number of crossroads.    
   
\subsection{Solving Strategy}\label{sec:solve}

This paper focuses on the presentation of the embedding problem as an ILP. 
We did not implement an algorithm, yet, which exploits the branching procedure as described in \refSec{complexity}.  
It is not straightforward how the decision tree could be reduced at certain stages.
As we do not consider the ratio of broken vertices to be fixed here, there is still an exponential overhead in the number of final simplified problems.
Even the simplified heuristic version is still a hard optimization problem. 

Thus using an ILP solver, already taking advantage of implemented branch-and-bound techniques, is a good starting point to evaluate the capabilities of the model.  
We decided to pass the models~\eqref{eq:EILP} and~\eqref{eq:HILP} directly to the solver SCIP~\cite{scip} without further adjustments.
It is currently one of the fastest non-commercial solvers for mixed integer programming, which includes ILP.

\subsection{Specifications}\label{sec:sepcifications}
    
The experiments were run on a Dell Precision 5820 Tower workstation with a Intel Xeon(R) W-2175 CPU @ 2.50GHz × 28, 128 RAM and operating system Ubuntu Linux 18.04.5 LTS.
We implemented our code in python and used the python interface package \texttt{pyscipopt}~\cite{pyscipopt} to connect to the solver SCIP with version 6.0.1~\cite{scip}.  
As this interface does not support parallel mode, we could just use one core.    
We set a timeout of 1 hour for solving each instance with SCIP.
Building up the model was not included in there. 
As we first want to evaluate the capabilities of the model and the derived heuristic version themselves, we did not optimize our code regarding performance.
Apart from the timeout we used the default SCIP parameters.

\section{Results}\label{sec:results}

A good reference to estimate the quality of a solution is to compare its objective value, the found graph size, to the largest possible size of a complete graph in the ideal Chimera graph. 
As stated in \refSec{approach} with our construction using crosses the largest complete graph in a Chimera graph of size $s$ is $K_{4s}$.
Let $G_{s, b, i}$ be the graph size returned by SCIP within one hour for the $i$'th instance with Chimera size $s$ and ratio of broken vertices $b$.  
As we consider 10 instances for each parameter combination the found graph sizes are averaged and 
we introduce the \emph{averaged solution ratio}
\begin{equation}
    \closure{R}_{s, b} := \frac{\frac{1}{10}\sum_{i = 0}^9 G_{s, b, i}}{4s} = \sum_{i = 0}^9 \frac{G_{s, b, i}}{40 s}
\end{equation}
as a measure for the solution quality.
\refTable{ratios} shows the resulting ratios for each of the models.

\pgfplotstableset{%
    column type=r,
    assign column name/.style={
        /pgfplots/table/column name={\textbf{#1}~~}
    },
    every head row/.style={after row=\hline},
    every column/.style={
        column type/.add={>{}}{<{\hspace{-0.5em}}},
        fixed, 
        fixed zerofill,
        precision=2, 
        clear infinite,
        empty cells with={\cellcolor[HTML]{D8D8D8}--~~},
    },
    every first column/.style={
        fixed,
        zerofill=false,
        precision=3,
        column type=>{\!}l<{\!\!}|,
        column name={\diagbox[innerwidth=2em]{$\bm{b}$}{\raisebox{-2mm}{$\bm{s}$}}},
        postproc cell content/.style={@cell content=$\bm{##1}$}
    },
}

\newcommand{\colorzerowhite}[1]{%
    \pgfmathfloatparsenumber{#1}%
    \pgfmathfloatifflags{\pgfmathresult}{0}{%
        \pgfplotstableset{/pgfplots/table/@cell content={\cellcolor[HTML]{FFFFFF}#1}}%
    }{}%
}

\newcommand{\colorcellsfromrow}[2]{%
    \ifnum\pgfplotstablerow>#1%
        \pgfplotstableset{/pgfplots/table/@cell content/.add={\cellcolor[HTML]{#2}}{}}%
    \else%
        \ifnum\pgfplotstablerow=#1%
            \pgfplotstableset{/pgfplots/table/@cell content/.add={\cellcolor[HTML]{#2}}{}}%
        \fi%
    \fi%
}

\begin{table}[p]
    \centering
    \scriptsize
    \subfloat[for exact model~\eqref{eq:EILP} \label{table:ratios:exact}]{\begin{filecontents*}{data/ratios_exact.txt}
ratio   4       6       8       10      12      14      16      18      20      22      24      26      28      30      32      34
0.005   1.00000 1.00000 1.00000 1.00000 1.00000 1.00000 1.00000 1.00000 1.00000 1.00000 1.00000 1.00000 1.00000 1.00000 1.00000 1.00000
0.01    1.00000 1.00000 1.00000 1.00000 1.00000 1.00000 1.00000 1.00000 1.00000 1.00000 1.00000 1.00000 0.96786 0.90667 0.82187 0.85368
0.02    1.00000 1.00000 1.00000 1.00000 1.00000 1.00000 0.99844 0.99861 0.99375 0.95114 0.77396 0.68173 0.69821 0.62750 0.60156 0.60441
0.03    1.00000 0.99583 0.99688 1.00000 1.00000 1.00000 1.00000 0.94167 0.62875 0.59432 0.59167 0.57788 0.51071 0.50750 0.52031 0.44265
0.04    1.00000 0.99583 0.99688 0.99250 0.98750 0.98750 0.88438 0.60139 0.60750 0.56023 0.50104 0.52115 0.43036 0.44333 0.42969 0.39265
0.05    0.98750 0.98750 0.98125 0.99000 0.97917 0.94286 0.64531 0.58750 0.42625 0.51136 0.43646 0.42692 0.37768 0.37000 0.34609 0.37647
0.1     0.95000 0.92500 0.85625 0.81000 0.69375 0.51786 0.36719 0.35000 0.30000 0.28636 0.25313 0.24712 0.23929 0.22417 0.21016 0.17206
0.2     0.71875 0.60000 0.52500 0.47750 0.38958 0.21964 0.20313 0.16389 0.16875 0.16705 0.12292 0.13750 0.10893 0.10417 nan     nan
\end{filecontents*}
\pgfplotstabletypeset[%
    columns/10/.style={postproc cell content/.append code={\colorcellsfromrow{6}{FFFFB4}}},
    every row 4 column 14/.style={postproc cell content/.append style={@cell content/.add={\cellcolor[HTML]{FFFFB4}}{}}},
    every row 5 column 14/.style={postproc cell content/.append style={@cell content/.add={\cellcolor[HTML]{FFFFB4}}{}}},
    every row 4 column 16/.style={postproc cell content/.append style={@cell content/.add={\cellcolor[HTML]{FFFFB4}}{}}},
    every row 3 column 18/.style={postproc cell content/.append style={@cell content/.add={\cellcolor[HTML]{FFFFB4}}{}}},
    every row 2 column 20/.style={postproc cell content/.append style={@cell content/.add={\cellcolor[HTML]{FFFFB4}}{}}},
    every row 2 column 22/.style={postproc cell content/.append style={@cell content/.add={\cellcolor[HTML]{FFFFB4}}{}}},
    every row 2 column 24/.style={postproc cell content/.append style={@cell content/.add={\cellcolor[HTML]{FFFFB4}}{}}},
    every row 1 column 28/.style={postproc cell content/.append style={@cell content/.add={\cellcolor[HTML]{FFFFB4}}{}}},
    every row 1 column 30/.style={postproc cell content/.append style={@cell content/.add={\cellcolor[HTML]{FFFFB4}}{}}},
    columns/12/.style={postproc cell content/.append code={\colorcellsfromrow{6}{FFCCC9}}},
    columns/14/.style={postproc cell content/.append code={\colorcellsfromrow{6}{FFCCC9}}},
    columns/16/.style={postproc cell content/.append code={\colorcellsfromrow{5}{FFCCC9}}},
    columns/18/.style={postproc cell content/.append code={\colorcellsfromrow{4}{FFCCC9}}},
    columns/20/.style={postproc cell content/.append code={\colorcellsfromrow{3}{FFCCC9}}},
    columns/22/.style={postproc cell content/.append code={\colorcellsfromrow{3}{FFCCC9}}},
    columns/24/.style={postproc cell content/.append code={\colorcellsfromrow{3}{FFCCC9}}},
    columns/26/.style={postproc cell content/.append code={\colorcellsfromrow{2}{FFCCC9}}},
    columns/28/.style={postproc cell content/.append code={\colorcellsfromrow{2}{FFCCC9}}},
    columns/30/.style={postproc cell content/.append code={\colorcellsfromrow{2}{FFCCC9}}},
    columns/32/.style={postproc cell content/.append code={\colorcellsfromrow{1}{FFCCC9}}},
    columns/34/.style={postproc cell content/.append code={\colorcellsfromrow{1}{FFCCC9}}},
]{data/ratios_exact.txt}
} \\
    \subfloat[for heuristic model~\eqref{eq:HILP} with $m=0.25$ \label{table:ratios:heur025}]{\begin{filecontents*}{data/ratios_heur_025.txt}
ratio   4       6       8       10      12      14      16      18      20      22      24      26      28      30      32      34
0.005   1.00000 1.00000 1.00000 1.00000 1.00000 1.00000 1.00000 1.00000 1.00000 1.00000 1.00000 1.00000 1.00000 1.00000 1.00000 1.00000
0.01    1.00000 1.00000 1.00000 1.00000 1.00000 1.00000 1.00000 1.00000 1.00000 1.00000 1.00000 1.00000 0.98750 0.88000 0.86719 0.86324
0.02    1.00000 1.00000 1.00000 1.00000 1.00000 1.00000 0.99844 0.99861 1.00000 0.94318 0.82604 0.75192 0.75625 0.76500 0.72656 0.68824
0.03    1.00000 0.99583 0.99688 1.00000 1.00000 1.00000 1.00000 0.99167 0.78125 0.69659 0.67292 0.67404 0.61875 0.61750 0.60781 0.59265
0.04    1.00000 0.99583 0.99688 0.99250 0.98750 0.98750 0.97344 0.74306 0.70250 0.63068 0.61354 0.60096 0.55982 0.52750 0.49453 0.48162
0.05    0.98750 0.98750 0.98125 0.99000 0.97708 0.95179 0.71563 0.66528 0.60250 0.57159 0.51563 0.50192 0.48839 0.45000 0.42188 0.41324
0.1     0.95000 0.92500 0.85625 0.81000 0.70833 0.58571 0.42813 0.40694 0.36250 0.32045 0.30729 0.29423 0.27589 0.23417 0.23906 0.21250
0.2     0.71875 0.60000 0.51563 0.47500 0.42083 0.35536 0.31250 0.21944 0.17000 0.15795 0.12813 0.12404 0.10893 0.09250 0.09219 0.07794
\end{filecontents*}
\pgfplotstabletypeset[%
    every row 6 column 10/.style={postproc cell content/.append style={@cell content/.add={\cellcolor[HTML]{FFFFB4}}{}}},
    every row 5 column 12/.style={postproc cell content/.append style={@cell content/.add={\cellcolor[HTML]{FFFFB4}}{}}},
    every row 6 column 12/.style={postproc cell content/.append style={@cell content/.add={\cellcolor[HTML]{FFCCC9}}{}}},
    every row 7 column 12/.style={postproc cell content/.append style={@cell content/.add={\cellcolor[HTML]{FFFFB4}}{}}},
    every row 4 column 14/.style={postproc cell content/.append style={@cell content/.add={\cellcolor[HTML]{FFFFB4}}{}}},
    every row 5 column 14/.style={postproc cell content/.append style={@cell content/.add={\cellcolor[HTML]{FFFFB4}}{}}},
    every row 4 column 16/.style={postproc cell content/.append style={@cell content/.add={\cellcolor[HTML]{FFFFB4}}{}}},
    every row 3 column 18/.style={postproc cell content/.append style={@cell content/.add={\cellcolor[HTML]{FFFFB4}}{}}},
    every row 2 column 22/.style={postproc cell content/.append style={@cell content/.add={\cellcolor[HTML]{FFFFB4}}{}}},
    every row 1 column 28/.style={postproc cell content/.append style={@cell content/.add={\cellcolor[HTML]{FFFFB4}}{}}},
    every row 1 column 30/.style={postproc cell content/.append style={@cell content/.add={\cellcolor[HTML]{FFFFB4}}{}}},
    columns/14/.style={postproc cell content/.append code={\colorcellsfromrow{6}{FFCCC9}}},
    columns/16/.style={postproc cell content/.append code={\colorcellsfromrow{5}{FFCCC9}}},
    columns/18/.style={postproc cell content/.append code={\colorcellsfromrow{4}{FFCCC9}}},
    columns/20/.style={postproc cell content/.append code={\colorcellsfromrow{3}{FFCCC9}}},
    columns/22/.style={postproc cell content/.append code={\colorcellsfromrow{3}{FFCCC9}}},
    columns/24/.style={postproc cell content/.append code={\colorcellsfromrow{2}{FFCCC9}}},
    columns/26/.style={postproc cell content/.append code={\colorcellsfromrow{2}{FFCCC9}}},
    columns/28/.style={postproc cell content/.append code={\colorcellsfromrow{2}{FFCCC9}}},
    columns/30/.style={postproc cell content/.append code={\colorcellsfromrow{2}{FFCCC9}}},
    columns/32/.style={postproc cell content/.append code={\colorcellsfromrow{1}{FFCCC9}}},
    columns/34/.style={postproc cell content/.append code={\colorcellsfromrow{1}{FFCCC9}}},
]{data/ratios_heur_025.txt}}   \\
    \subfloat[for heuristic model~\eqref{eq:HILP} with $m=0.0$ \label{table:ratios:heur}]{\begin{filecontents*}{data/ratios_heur_0.txt}
ratio   4       6       8       10      12      14      16      18      20      22      24      26      28      30      32      34
0.005   1.00000 1.00000 1.00000 1.00000 1.00000 1.00000 1.00000 1.00000 1.00000 1.00000 1.00000 1.00000 1.00000 1.00000 1.00000 1.00000
0.01    1.00000 1.00000 1.00000 1.00000 1.00000 1.00000 1.00000 1.00000 1.00000 1.00000 1.00000 1.00000 1.00000 1.00000 1.00000 1.00000
0.02    1.00000 1.00000 1.00000 1.00000 1.00000 1.00000 0.99844 0.99861 1.00000 1.00000 1.00000 1.00000 0.99464 0.97917 0.97188 0.92426
0.03    1.00000 0.99583 0.99688 1.00000 1.00000 1.00000 0.99688 0.98750 0.96250 0.93523 0.89063 0.85962 0.80804 0.78500 0.74219 0.68676
0.04    1.00000 0.99583 0.99688 0.99250 0.98542 0.98036 0.95781 0.90278 0.84000 0.78636 0.74583 0.68365 0.61696 0.57833 0.52969 0.48971
0.05    0.98750 0.98750 0.98125 0.99000 0.96250 0.90893 0.82031 0.77500 0.68375 0.62841 0.56771 0.50288 0.45357 0.39250 0.37266 0.35221
0.1     0.95000 0.89583 0.76875 0.66500 0.50833 0.42857 0.36875 0.28750 0.25750 0.18636 0.15104 0.13942 0.11161 0.08000 0.05859 0.05662
0.2     0.66250 0.42500 0.28438 0.22000 0.13125 0.09107 0.06406 0.02917 0.02000 0.01477 0.00729 0.00769 0.00446 0.00500 0.00078 0.00000
\end{filecontents*}
\pgfplotstabletypeset[%
    every row no 5 column no 7/.style={postproc cell content/.append style={@cell content/.add={\cellcolor[HTML]{FFFFB4}}{}}},
    every row 4 column 18/.style={postproc cell content/.append style={@cell content/.add={\cellcolor[HTML]{FFFFB4}}{}}},
    every row 3 column 20/.style={postproc cell content/.append style={@cell content/.add={\cellcolor[HTML]{FFFFB4}}{}}},
    every row 4 column 20/.style={postproc cell content/.append style={@cell content/.add={\cellcolor[HTML]{FFFFB4}}{}}},
    every row 5 column 20/.style={postproc cell content/.append style={@cell content/.add={\cellcolor[HTML]{FFFFB4}}{}}},
    every row 3 column 22/.style={postproc cell content/.append style={@cell content/.add={\cellcolor[HTML]{FFFFB4}}{}}},
    every row 4 column 22/.style={postproc cell content/.append style={@cell content/.add={\cellcolor[HTML]{FFFFB4}}{}}},
    every row 3 column 24/.style={postproc cell content/.append style={@cell content/.add={\cellcolor[HTML]{FFFFB4}}{}}},
    every row 4 column 24/.style={postproc cell content/.append style={@cell content/.add={\cellcolor[HTML]{FFFFB4}}{}}},
    every row 3 column 26/.style={postproc cell content/.append style={@cell content/.add={\cellcolor[HTML]{FFFFB4}}{}}},
    every row 4 column 26/.style={postproc cell content/.append style={@cell content/.add={\cellcolor[HTML]{FFFFB4}}{}}},
    every row 2 column 28/.style={postproc cell content/.append style={@cell content/.add={\cellcolor[HTML]{FFFFB4}}{}}},
    every row 3 column 28/.style={postproc cell content/.append style={@cell content/.add={\cellcolor[HTML]{FFFFB4}}{}}},
    every row 2 column 30/.style={postproc cell content/.append style={@cell content/.add={\cellcolor[HTML]{FFFFB4}}{}}},
    every row 3 column 30/.style={postproc cell content/.append style={@cell content/.add={\cellcolor[HTML]{FFFFB4}}{}}},
    every row 4 column 30/.style={postproc cell content/.append style={@cell content/.add={\cellcolor[HTML]{FFFFB4}}{}}},
    every row 2 column 32/.style={postproc cell content/.append style={@cell content/.add={\cellcolor[HTML]{FFFFB4}}{}}},
    every row 3 column 32/.style={postproc cell content/.append style={@cell content/.add={\cellcolor[HTML]{FFFFB4}}{}}},
    every row 4 column 32/.style={postproc cell content/.append style={@cell content/.add={\cellcolor[HTML]{FFFFB4}}{}}},
    every row 2 column 34/.style={postproc cell content/.append style={@cell content/.add={\cellcolor[HTML]{FFCCC9}}{}}},
    every row 3 column 34/.style={postproc cell content/.append style={@cell content/.add={\cellcolor[HTML]{FFFFB4}}{}}},
    every row 4 column 34/.style={postproc cell content/.append style={@cell content/.add={\cellcolor[HTML]{FFFFB4}}{}}},
]{data/ratios_heur_0.txt}
}
    \caption{Averaged solution ratio $\closure{R}_{s, b}$ for each combination of size $s$ and ratio of broken vertices $b$.
             Colors indicate the number of instances that where solved to optimality (white: all 10, yellow: 1 to 9, red: none).} 
    \label{table:ratios}         
    \vspace{0.5cm}
    \begin{filecontents*}{data/diff_ratios_heurs.txt}
ratio   4     6     8     10    12    14    16    18    20    22    24    26    28    30    32    34
0.005   0.00  0.00  0.00  0.00  0.00  0.00  0.00  0.00  0.00  0.00  0.00  0.00  0.00  0.00  0.00  0.00     
0.01    0.00  0.00  0.00  0.00  0.00  0.00  0.00  0.00  0.00  0.00  0.00  0.00 -0.01 -0.12 -0.13 -0.14     
0.02    0.00  0.00  0.00  0.00  0.00  0.00  0.00  0.00  0.00 -0.06 -0.17 -0.25 -0.23 -0.21 -0.24 -0.23     
0.03    0.00  0.00  0.00  0.00  0.00  0.00  0.00  0.00 -0.18 -0.24 -0.22 -0.19 -0.19 -0.17 -0.08 -0.10     
0.04    0.00  0.00  0.00  0.00  0.00  0.01  0.01 -0.16 -0.14 -0.16 -0.14 -0.08 -0.06 -0.05 -0.04 -0.01     
0.05    0.00  0.00  0.00  0.00  0.02  0.04 -0.10 -0.11 -0.08 -0.06 -0.05  0.00  0.04  0.06  0.05  0.06      
0.1     0.00  0.03  0.09  0.14  0.20  0.16  0.06  0.12  0.10  0.13  0.16  0.15  0.17  0.15  0.18  0.15      
0.2     0.06  0.17  0.24  0.25  0.29  0.27  0.25  0.19  0.15  0.15  0.12  0.11  0.11  0.08  0.09  0.08
\end{filecontents*}
\pgfplotstabletypeset[%
    fonts by sign={\cellcolor[HTML]{DAFFB4}}{\hspace{-1em}\cellcolor[HTML]{CAF4FE}},%
    every column/.append style={postproc cell content/.append code={\colorzerowhite{##1}}},%
]{data/diff_ratios_heurs.txt}

    \caption{Difference of rounded averaged solution ratios from heuristic models. Colors indicate advantage of a certain model (green: $m=0.25$, blue: $m=0.0$).}
    \label{table:diff}
\end{table}

In \refTable{ratios:exact} we can see a clear boundary between the instances which can be solved in one hour and which cannot.
The former are either instances with a small Chimera size or with a small ratio of broken vertices. 
Here we already see a slight advantage in the achieved graph sizes over~\cite{boothby2016fast}. 

Besides, we also solved the exact model for both versions of the aforementioned D-Wave 2000Q chips with 7 and 17 broken vertices, respectively. 
Not surprisingly in accordance to the results of \refTable{ratios:exact}, 
we were able to find an embedding for the complete graph with 64 vertices in both cases. 

For the unsolved instances we use the current best solution SCIP provides at the timeout, being a proven lower bound on the actual optimum. 
As SCIP is a MIP solver it tries to solve a given model to proven optimality and thus is not made for calculating fast approximate solutions. 
Therefore the found solutions for instances with increasing Chimera sizes and ratios decrease significantly, due to the sizes of the models. 
The instance combinations $(32, 0.2)$ and $(34, 0.2)$ could not be solved at all with the exact model because SCIP ran out of memory.
Nevertheless, the remaining instances provide values comparable to those of~\cite{boothby2016fast}.

Evaluating the heuristic approach, we can see that the \refTable{ratios:heur025} for $m=0.25$ does not differ significantly from the one for the exact model according to solvability.
Having a closer look to the values, we see no decrease for the solved instances.  
However, there is a slight improvement for the unsolved instances. 
Thus the model has a slight advantage regarding runtime but still seems to yield close to optimal values. 

Regarding the heuristic approach with $m=0$, shown in \refTable{ratios:heur}, we have a much stronger difference to the exact model. 
A lot more instances could be solved within one hour of computation time, especially those with larger ratios of broken vertices.
For the combination of sizes above 18 and ratios between about 0.02 and 0.05 we see a clear improvement through the heuristic, although a few instances could not be solved, too. 
This region of parameters, where it is reasonable to use this heuristic, is also clearly recognizable in \refTable{diff}, 
where we show the advantage of the heuristics with either $m=0$ or $m=0.25$.

However, in \refTable{ratios:heur} we observe that the proportions of graph sizes are much smaller for ratios above 0.1 than for $m=0.25$.    
Thus this heuristic model seems to get easier again with an increasing ratio of broken vertices.  
We assume a significant number of crossroads is excluded in advance, because of the large number of broken vertices, 
such that the resulting model has only very few solutions left.
In these cases the heuristic with $m=0$ is much too restrictive and $m=0.25$ is advantageous.   

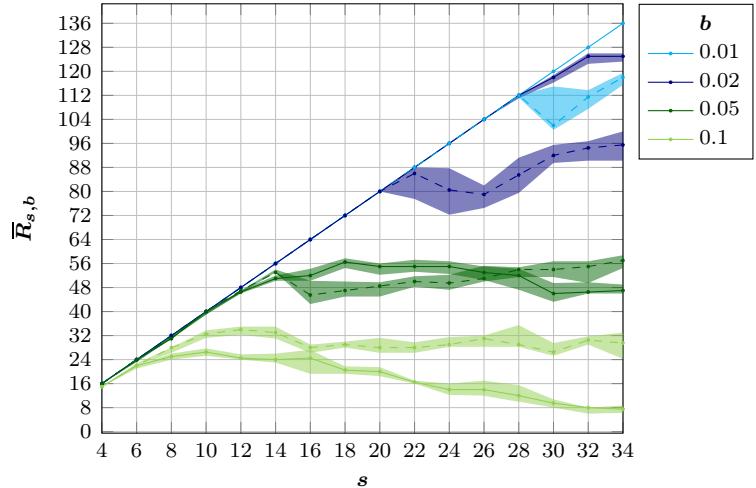
\begin{figure}[t]
    \centering

\newcommand{\plotData}[8]{

	\addplot+ [
		x filter/.expression={\thisrow{#1}==#8 ? x : nan},
		no markers, 
		draw=none, 
		name path=min,
		forget plot
	]
		table[
			x=#2,
			y=#5
		] {#6};
		
	\addplot+ [
		x filter/.expression={\thisrow{#1}==#8 ? x : nan},
		no markers, 
		draw=none,
		name path=max,
		forget plot
	]
		table[
			x=#2,
			y=#4
		] {#6};

	\addplot+[opacity=0.5, forget plot, on layer={}] fill between [of=min and max];
	
	\addplot+ [
		x filter/.expression={\thisrow{#1}==#8 ? x : nan},
		mark size=0.5pt, 
		mark=*, 
		#7
	]
		table[
			x=#2,
			y=#3
		] {#6};
}

\pgfplotscreateplotcyclelist{mylist}{
	{lightblue},
	{darkblue},
	{darkgreen},
	{lightgreen},
}

\begin{externalize}
\begin{tikzpicture}
	
	\pgfplotstableread{data/plot_heur_0.txt}\loadedtable
	\pgfplotstableread{data/plot_heur_025.txt}\loadedtabletwo
	\begin{axis}[
		xlabel={$\bm{s}$},
		ylabel={$\closure{\bm{R}}_{\bm{s, b}}$},
		scaled ticks=false,
		ytick distance=8,
		xtick distance=2,
	    enlarge x limits=false, 
	    enlarge y limits=0.05, 
	    cycle list name=mylist,	
	    legend cell align=left,
		legend pos=outer north east,
		grid=both,
        grid style={line width=.1pt, draw=gray!10},
        major grid style={line width=.2pt,draw=gray!50},
	]

        \addlegendimage{empty legend}
        \addlegendentry{$\bm{b}$}
        
		\foreach \val in {0.01, 0.02, 0.05, 0.1}{
			\plotData{ratio}{size}{Median}{Q14}{Q34}{\loadedtable}{}{\val}
		}	
		\addlegendentry{0.01}
        \addlegendentry{0.02}
        \addlegendentry{0.05}
        \addlegendentry{0.1}
	
        \pgfplotsset{cycle list shift=1}
        \foreach \val in {0.01, 0.02, 0.05, 0.1}{
            \plotData{ratio}{size}{Median}{Q14}{Q34}{\loadedtabletwo}{dashed}{\val}
        }   
	\end{axis}
\end{tikzpicture}
\end{externalize}
    \caption{Complete graph sizes against Chimera sizes $s$ for selected ratios of broken vertices $b$ for our heuristic approach~\eqref{eq:HILP}.
             The median of $G_{s,b,i}$ over $i \in [10]$ is depicted by the solid lines for $m=0.0$ and dashed lines for $m=0.25$, the shaped regions illustrate the quartiles.}
    \label{fig:plot}
\end{figure}

In order to compare our approach to previous work by Boothby~et.al.~\cite{boothby2016fast} we plot the found graph sizes for selected ratios of broken vertices in \refFig{plot}.
For larger ratios of broken qubits, e.g.\ $b=0.1$ and $b=0.05$, the maximum over the found graph sizes is comparable to~\cite{boothby2016fast} for both $m=0.0$ and $m=0.25$.
In contrast, for smaller ratios of broken qubits, e.g.\ $b=0.01$ and $b=0.02$ our heuristic approach with $m=0.0$ is able to embed larger complete graphs than it was reported in~\cite{boothby2016fast}.  
Note, that the diagonal corresponds to representing the largest possible complete graph size.       

All in all, for a ratio of 0.05 or smaller we observe that the proportions from the solved instances with small size and ratio have a value very close to 1.0, 
meaning most of them yield a maximal or close to maximal complete graph despite the presence of broken vertices.
Due to the heuristic results, we expect just a very small decline in the proportions for the exact model for larger Chimera sizes, 
if we could solve them to the end. 
This is based on the fact that the heuristics provide a lower bound on the actual optimum of the exact model. 
Thus despite the shortcomings presented in \refSec{non-optimal} our model is indeed very powerful.  

\section{Conclusion}\label{sec:conclusion}

We introduced a novel approach for the problem of embedding a complete graph into a faulty Chimera hardware architecture. 
It is based on a formulation as a bipartite matching optimization problem with additional constraints. 
We could show by a detailed analysis that the problem is fixed parameter tractable, where the decisive parameter is the number of broken vertices. 
The formulated optimization problem~\eqref{eq:EILP} can be solved to optimality using state-of-the-art MIP solvers for small Chimera sizes or a small ratio of broken vertices. 
Especially in these parameter settings the optimal value of the heuristic version~\eqref{eq:HILP} does not differ significantly from the original one. 
For larger Chimera graphs with a ratio of broken vertices in a certain range the heuristic performs even better within the given time constraint of one hour, 
due to the removal of unlikely crossroads and thus several constraints. 
However, if the ratio is too large, here above 0.1, the heuristic is too restrictive and the solution quality decreases again. 
Nevertheless the complete graph sizes we have found exceed the ones from previous approaches~\cite{klymko2014adiabatic,boothby2016fast}.  
 
Further regarding the current developments in the area of quantum annealers, larger graphs with less broken vertices,
and operational times of over one year we can produce reusable templates for complete graphs with a reasonable computational power.
Nevertheless there is some space for improvements. 
By exploiting the full potential of the branching strategy shown in \refSec{complexity} 
using dedicated bounding techniques we could develop a more customized, exact or heuristic, solver.  

Another step is to transfer the approach to the just recently released new hardware graph Pegasus. 
It yields a larger connectivity for the same number of vertices, but at the same time this makes the Pegasus graph less approachable.
The shown constructions for the Chimera graph provide a deeper insight into the structure of such lattice-like graphs and the problems dealing with them.  
Observing the physical realization by specifically arranged overlapping loops one can see that the Pegasus graph is closely related to the Chimera~\cite{dwavedocs}. 
Thus we are confident that our model construction for the Chimera can be transferred to the Pegasus topology.
Due to the larger vertex degree we even expect less constraints resulting from broken vertex pairs than for the Chimera.


\paragraph{Acknowledgement} 
    The authors gratefully acknowledge the Jülich Supercomputing Centre (JSC) for supporting this work 
    by providing the access to the D-Wave 2000Q through the Jülich UNified Infrastructure for Quantum computing (JUNIQ).


    \bibliographystyle{abbrv}
    \bibliography{references}

\end{document}